\documentclass[letterpaper,journal]{IEEEtran}

\usepackage{etoolbox}
\providetoggle{original}
\settoggle{original}{false}
\usepackage[utf8]{inputenc} 
\usepackage[T1]{fontenc}
\usepackage{newtxtext}
\usepackage{newtxmath}
\usepackage{citesort}

\usepackage{ragged2e}
\usepackage{amsmath,amsfonts}
\usepackage{algorithm}
\usepackage{algpseudocode}
\algnewcommand{\INPUT}{\item[\textbf{Input:}]}
\algnewcommand{\OUTPUT}{\item[\textbf{Output:}]}
\usepackage{array}
\usepackage[caption=false,font=scriptsize,labelfont=rm,textfont=rm]{subfig}
\usepackage{textcomp}
\usepackage{stfloats}
\usepackage{url}
\usepackage{verbatim}
\usepackage{graphicx}
\usepackage{soul}

\usepackage{xcolor}
\definecolor{tempbyliaw}{rgb}{0,0.3,.6}
\definecolor{myblue}{rgb}{1, 0.75, 0.0}   

\usepackage{booktabs}
\usepackage{tabularx}
\usepackage{makecell}

\begin{document}

\title{MemAscend: System Memory Optimization for SSD-Offloaded LLM Fine-Tuning}

\author{
    Yong-Cheng Liaw, 
	Shuo-Han Chen,~\IEEEmembership{Member,~IEEE}\vspace{-0.2in}
}

\maketitle

\begin{abstract}\justifying
Owing to the huge success of generative artificial intelligence (AI), large language models (LLMs) have emerged as a core subclass, powering applications such as question answering, text generation, and code completion. While fine-tuning these models on domain-specific data can yield significant performance gains, it also poses daunting computational challenges, especially for researchers and organizations with limited hardware resources. Although SSD offloading (i.e., ZeRO-Infinity) has emerged as a viable strategy to overcome the GPU memory barrier via leveraging both system memory (i.e., CPU DRAM) and storage space (i.e., solid-state devices, SSDs), its design primarily targets model-centric performance issues. As a result, key system-level issues, including system memory fragmentation, inefficient pinned buffer allocation, peak system memory usage spikes, and file system overhead, remain unaddressed, limiting scalability and inflating costs. Such an observation motivates this paper to introduce MemAscend, a framework that systematically tackles the underexplored system memory bottlenecks in SSD-offloaded LLM training, with a focus on resource-constrained environments. By streamlining pinned-memory allocation, eradicating fragmentation, and mitigating peak system memory usage, MemAscend reclaims a substantial system memory budget, enabling larger models, longer context windows, and higher batch sizes without exceeding modest hardware limits. Across diverse LLM benchmarks, MemAscend reduces peak system-memory consumption by an average of 55.7\% compared with state-of-the-art SSD offloading techniques, lowering the hardware barrier for fine-tuning and unlocking new possibilities for cost-effective large-scale training on limited-resource machines.
\end{abstract}


\IEEEpeerreviewmaketitle

\vspace{-0.3in}
\section{Introduction}
\label{sec:introduction}
The growth of large-language-model (LLM) applications has significantly surged in recent years, propelled by steady advances in model capabilities~\cite{minaee2025largelanguagemodelssurvey}. LLMs have now powered conversational assistants, automatic summarization, multilingual translation, and countless other tasks~\cite{openai2025gpt45systemcard,grattafiori2024llama3herdmodels,chhibbar2024automatic,hemamou2024scalingupsummarization,cui2025multilingualmt}. To adapt LLMs to a specific domain, fine-tuning on domain-specific datasets is a widely used approach~\cite{zhang2024instructiontuninglargelanguage} to customize pre-trained foundational models and address the nuances of a specific domain better. Although fine-tuning markedly boosts reliability and task performance, it also demands considerable computation and memory usage. Commercial leaders routinely train foundation models on massive GPU clusters. For example, Meta’s Llama 3 relied on 16,000 NVIDIA H100 GPUs \cite{grattafiori2024llama3herdmodels}. By contrast, academic labs, individual researchers, and small enterprises are often operating within tight hardware budgets, with limited GPU count and modest system-memory capacity. Finding cost-effective ways to fine-tune LLMs under such tight resource constraints, therefore, remains an open challenge.


Broadly, LLM training consumes two memory classes: (1) \textbf{static memory}, which includes model weights and optimizer states, and (2) \textbf{residual states}, such as activations and intermediate values~\cite{rajbhandari2020zeromemoryoptimizationstraining}. For residual states, activations dominate due to their linear growth with batch size and sequence length~\cite{Li2022harmony}, and several techniques have been proposed to mitigate the above memory pressure. For instance, to lower activation memory, gradient checkpointing~\cite{chen2016trainingdeepnetssublinear} is designed not to store every activation during the forward pass, but instead stores only a few checkpointed activations and recomputes those missed ones during backpropagation. The intermediate values, which are temporary results that persist across multiple operations, such as across multiple CUDA kernels\footnote{CUDA (Compute Unified Device Architecture) kernels are functions, also basic execution units, written in CUDA C/C++ that run on the GPU.}~\cite{cudadocumentation}, significantly contribute to peak memory usage. To lower the size of intermediate values, methods like Liger-Kernel~\cite{hsu2025ligerkernelefficienttriton} fuse operations (e.g., RMSNorm, SwiGLU, Cross-Entropy~\cite{zhang2019rootmeansquarelayer,shazeer2020gluvariantsimprovetransformer,zhang2018generalizedcrossentropyloss}) into single Triton~\cite{tironslanguage} kernels, eliminating transient buffers and launch overhead. Additionally, memory-efficient kernels like Flash-Attention~\cite{dao2022flashattentionfastmemoryefficientexact} tile the computation to avoid materializing the entire attention matrix at once, thereby reducing the size of intermediate values. Other approaches, such as mixed-precision training~\cite{micikevicius2018mixedprecisiontraining} and related quantization techniques, can reduce the memory footprint of both activations and intermediate values by casting tensors to \texttt{fp16} or \texttt{bf16}.

On the other hand, static memory is substantial because modern transformers contain billions of parameters, and optimizers such as Adam store both momentum and variance for each weight~\cite{kingma2017adammethodstochasticoptimization}. To lower the amount of static memory, quantization has been widely explored. For instance, 8-bit optimizers \cite{dettmers20228bitoptimizersblockwisequantization} and 4-bit optimizers \cite{li2023memoryefficientoptimizers4bit} have shown promising results in reducing static memory consumption with minimal impact on convergence speed. On the other hand, Parameter-Efficient Fine-Tuning (PEFT) \cite{han2024parameterefficientfinetuninglargemodels}, especially Low-Rank Adaptation (LoRA) \cite{hu2021loralowrankadaptationlarge}, significantly reduces the number of trainable parameters by focusing on low-rank matrices while keeping the pre-trained model frozen. Although methods like LoRA offer substantial computational and memory efficiency, they often require meticulous hyperparameter tuning, such as determining the rank and selecting which weights to train, to achieve optimal performance \cite{biderman2024loralearnsforgets,shuttleworth2024loravsfinetuningillusion}. Consequently, full-parameter training remains crucial for maximizing task-specific performance, particularly in scenarios where pre-trained models have not been exposed to domain-specific or private data. In such cases, the discrepancies between the target and pre-trained domains can be substantial, necessitating full-parameter training to bridge the gap effectively.

Although full-parameter fine-tuning can achieve optimal performance in fine-tuning scenarios, the large size of pre-trained models limits the feasibility of training in resource-limited environments. Typically, fine-tuning tasks involve relatively smaller datasets, so the primary bottleneck in resource-limited scenarios is that GPU memory is insufficient to handle the static and residual memory requirements of these models. To address these limitations, SSD offloading has emerged as a promising solution. SSD offloading transfers static memory, such as model weights and optimizer states, from GPUs to SSDs, which generally provide significantly larger storage capacity. For example, ZeRO-Infinity~\cite{rajbhandari2021zeroinfinitybreakinggpumemory}, a state-of-the-art SSD offloading technique, moves static memory to NVMe SSDs. With SSD offloading, GPU memory is dedicated to storing intermediate values during computation.

System memory serves as an intermediary, either facilitating data transfers between the GPU and SSD or handling optimizer steps on the CPU to reduce data movement across the PCIe bus. Although SSD offloading may reduce training performance due to additional PCIe transfer latency, this impact is mitigated for fine-tuning tasks since fine-tuning tasks typically involve a relatively smaller dataset when compared with pre-trained tasks. Furthermore, SSD offloading, which targets alleviating the static memory bottleneck, can be integrated effectively with residual memory optimization techniques, such as Flash-Attention, Liger-Kernel, gradient checkpointing
, or further offloading the checkpointed gradient into system memory. By integrating these techniques, GPU memory usage is further lowered, enabling the largest trainable context length or batch size, and potentially mitigating the original performance loss caused by additional data transfers across the GPU, system memory, and SSDs.

However, existing SSD offloading techniques (i.e., ZeRO-Infinity) are primarily designed to address model-centric performance issues. While GPU memory usage has been extensively studied, the system memory usage has received comparatively less attention and, according to our investigation, the system memory usage has emerged as a new bottleneck in the SSD offloading scenario. In this study, key inefficiencies are found through analyzing system memory requirements for SSD offloading. For example, although pinned memory~\cite{cudapinnedmemory2012} (i.e., \texttt{cudaHostAlloc}) can be used to facilitate GPU-to-CPU data transfers, its current management with ZeRO-Infinity causes system memory fragmentation that wastes 72.71\% of buffer pool memory across different models, introducing unnecessary allocation overhead for system memory. Specifically, aggressive alignment in pinned memory allocation, while beneficial in data transfers, results in additional overhead and can double the allocation cost. Furthermore, inefficient CPU overflow checking contributes to peak memory overhead, leading to up to 1.25 times more memory usage depending on the size of the process tensor. In the end, our investigation reveals that these unnecessary system memory usage scales with model size, consistently causing an average of 55.7\% system memory waste across different model sizes. This inefficiency limits the trainable model size and the available space for offloaded checkpoints, thereby constraining the maximum context length and the ability to use larger batch sizes, leading to suboptimal performance due to insufficient batch size. Although several works have been proposed after ZeRO-Infinity, such as \cite{liao2024lohanlowcosthighperformancesystem,jang2024smartinfinityfastlargelanguage,wu2025ssdtrainactivationoffloadingsystem}, for further performance or efficiency enhancement, these approaches either focus on different aspects than ZeRO-Infinity or target specific scenarios, making them unsuitable for broader training use cases. As a result, the major unnecessary waste of system memory still remains unaddressed.

To address these challenges, this paper introduces MemAscend, a framework that aims to minimize system memory requirements within SSD offloaded fine-tuning while maximizing trainable model size and context length. MemAscend comprises four key components: (1) an adaptive buffer pool, (2) an alignment-free pinned memory allocation, (3) a fused overflow check mechanism, and (4) a direct NVMe engine. Firstly, the adaptive buffer pool eliminates memory fragmentation by adaptively fitting offloaded tensors into buffers on the system memory with the most suitable size. Secondly, the alignment-free pinned memory allocation avoided the alignment overhead between pinned memory allocations by implementing custom operations (i.e., c++ extensions) to avoid unnecessary alignment when allocating memory space. Thirdly, the fused overflow check mechanism is proposed to merge existing overflow-checking operations to eliminate the issue of double-peak memory usage. Lately, the direct NVMe engine bypasses the file system and enables direct data management on SSDs for low-latency SSD access. This study also demonstrates that the I/O transfer volume can be further reduced by integrating a pure half-precision optimizer, which transfers parameters, gradients, and momentum in half-precision at each optimizer step. Based on the above, the contributions of this study are summarized as follows:

\begin{enumerate}
    \item \textbf{Comprehensive system memory optimization}: MemAscend reduces peak system memory consumption by an average of 55.7\% through adaptive buffering, alignment-free pinned memory allocation, and removal of overflow-induced memory duplication.
    \item \textbf{Enhanced SSD I/O efficiency}: By integrating a Direct NVMe Engine and a half-precision optimizer, MemAscend reduces I/O transfer volume by 58\% and improves throughput by up to 24.21\% or 56.80\%, depending on the hardware configuration, while eliminating filesystem and staging overheads.
    \item \textbf{Improved scalability for fine-tuning scenario}: On a system with 128 GiB of memory, MemAscend enables scaling context lengths from 16,384 to 131,072 tokens or increasing batch size from 4 to 32 under the same configuration.
\end{enumerate}

The rest of this article is organized as follows. The background and motivation are detailed in Sections~\ref{sec:background} and~\ref{sec:observation}. The design of the proposed system is introduced in Section~\ref{sec:method}. Section~\ref{sec:analysis} provides analyses that address memory efficiency and explain how the proposed approach achieves the largest trainable model size and context length under identical configurations. The evaluation results are presented in Section~\ref{sec:evaluation}. Finally, Section~\ref{sec:conclusion} offers concluding remarks.
 
\vspace{-0.1in}
\section{Background and Related Works}
\label{sec:background}
This section provides background and related works by reviewing key developments in memory management for large-scale model training. It first outlines SSD offloading frameworks that extend GPU memory capacity using heterogeneous storage. It then discusses related SSD offloading techniques and their applicability across different training scenarios. Finally, it highlights GPU memory efficiency optimizations that complement offloading strategies.
 
\vspace{-0.15in}
\subsection{SSD offloading framework}
\label{subsec:ssd_offloading_systems}

\begin{figure}[b]
    \vspace{-0.2in}
    \centering
    \includegraphics[width=2.5in]{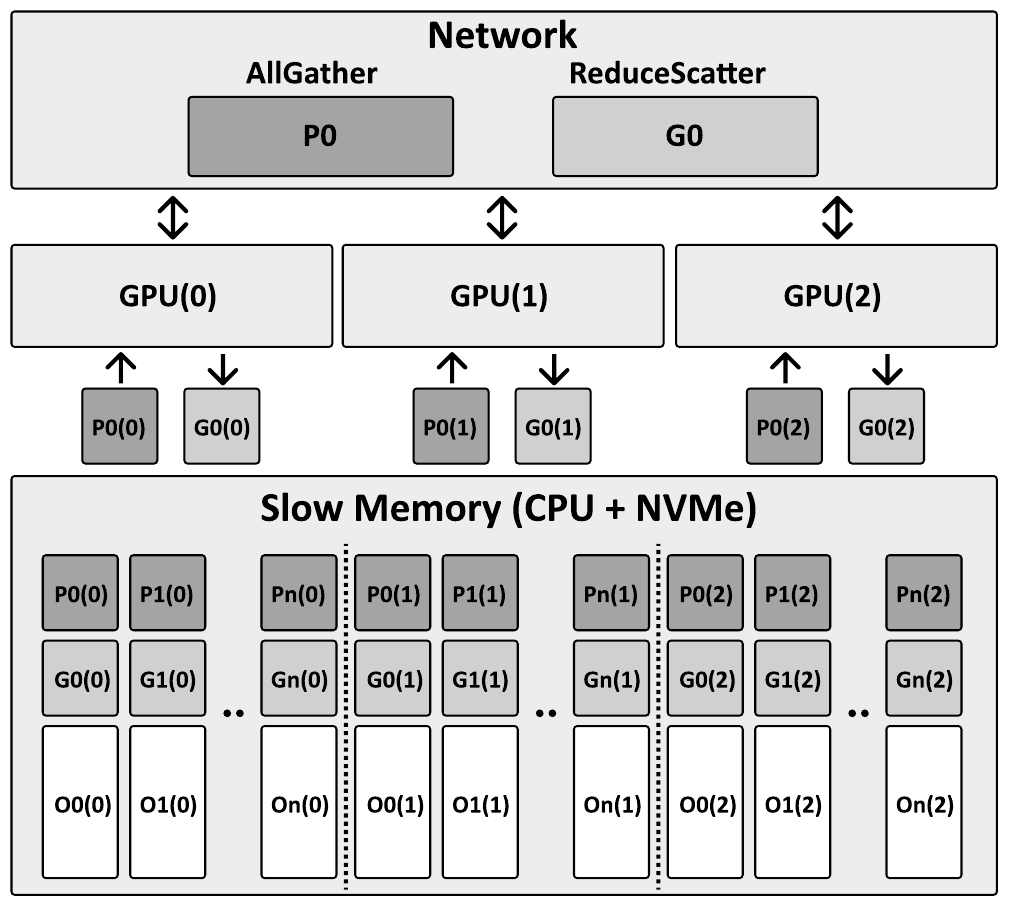} 
    \vspace{-0.1in}
    \caption{ZeRO-Infinity data flow during the backward pass for an $n$-layer model with three data-parallel ranks. In this example, network refers to the use of peer-to-peer communication operations, such as Allgather or ReduceScatter, which synchronize data (i.e., tensors) across multiple GPUs during distributed training. Parameters P0(i), which are the first tensors for the i-th GPU, are moved to each GPU. The gradients G0(i), which correspond to P0(i), are processed and offloaded. The optimizer states O0(i), which correspond to P0(i) and G0(i), are updated on the CPU.}
    \label{fig:zero_infinity_overview}
\end{figure}

To address GPU memory limitations in large-scale model training, heterogeneous memory systems that incorporate SSD offloading have emerged as a promising solution. These systems strategically relocate static model states, where tensor sizes vary dynamically due to mixed-precision storage formats, layer heterogeneity, and activation materialization boundaries. These tensors are not always needed during different stages of computation within a single training step, allowing them to be moved from expensive GPU memory to more abundant and cost-effective system memory or NVMe SSDs when not needed. The SSD acts as expanded but slower memory for storing data. Unlike traditional storage systems, where hot and cold data exhibit clear access-frequency disparities, SSD offloading during training accesses all offloaded tensors once per iteration. Therefore, all such data is effectively hot at every step. ZeRO-Infinity \cite{rajbhandari2021zeroinfinitybreakinggpumemory} is a state-of-the-art (SOTA) SSD offloading system integrated into the DeepSpeed library \cite{deepspeedai_deepspeed}. Building on the ZeRO distributed training strategy \cite{rajbhandari2020zeromemoryoptimizationstraining,ren2021zerooffloaddemocratizingbillionscalemodel}, which partitions model states across data-parallel processes to eliminate memory redundancy, ZeRO-Infinity extends the design by introducing an SSD offloading scheduler. This scheduler enables partitioned static model states to be offloaded first to CPU memory and subsequently to NVMe SSDs, as illustrated in Figure~\ref{fig:zero_infinity_overview}.

ZeRO-Infinity executes forward and backward passes on the GPU, but it typically offloads the optimizer step to the CPU. This is because the optimizer’s low arithmetic intensity rarely justifies the PCIe transfer costs required to move optimizer states back to the GPU~\cite{ren2021zerooffloaddemocratizingbillionscalemodel}. DeepSpeed supports this via a host optimizer with a fused C++/AVX backend. To maximize efficiency, this backend uses contiguous tensors and template dispatch for kernel selection. It supports fp16/bf16 parameters paired with either fp32 or low-precision optimizer states. For performance, updates are vectorized using AVX512/AVX2, while an OpenMP-parallel tiled loop handles remaining elements, including bias correction and decoupled weight decay. To optimize PCIe/NVMe throughput, ZeRO-Infinity employs bandwidth-centric partitioning and overlap-centric execution. By chunking tensors and staggering transfers, the system saturates NVMe bandwidth while asynchronously staging upcoming weights from the SSD to DRAM. However, relying on CPU memory as a staging buffer remains a bottleneck on systems with limited DRAM, a limitation this work aims to resolve.

\vspace{-0.1in}
\subsection{Related Works of SSD offloading}
\label{subsec:related_ssd_offloading}
Following the introduction of the state-of-the-art SSD offloading system, ZeRO-Infinity, subsequent studies have investigated various strategies within the broader context of SSD offloading. While some of these approaches, such as activation offloading, are compatible with ZeRO-Infinity and can be integrated to further enhance its performance, others are designed for specific hardware setups or constrained training environments, addressing distinct use cases that complement ZeRO-Infinity but do not directly generalize to its distributed training framework. This section reviews these related works in detail.


\subsubsection{SSDTrain}
SSDTrain~\cite{wu2025ssdtrainactivationoffloadingsystem} is an adaptive system designed to offload activations to NVMe SSDs, mitigating GPU memory bottlenecks caused by large activation tensors, particularly in training workloads with large batch sizes or long sequences. The system emphasizes efficient overlap between I/O transfers and GPU computation, achieving substantial reductions in peak activation memory usage with negligible performance overhead. SSDTrain's focus on activations complements systems like ZeRO-Infinity (and our MemAscend), which primarily offload static model states. Activation offloading techniques, such as those in SSDTrain, can potentially be integrated with model state offloading systems to further optimize end-to-end memory usage. As such, SSDTrain targets a different memory component and remains compatible with our offloading strategy.


\subsubsection{Smart-Infinity}
Smart-Infinity~\cite{jang2024smartinfinityfastlargelanguage} tackles the storage bandwidth bottleneck in SSD-offloaded LLM training by utilizing near-storage processing (NSP) hardware. It introduces SmartUpdate, a mechanism that offloads parameter updates to custom accelerators within computational storage devices (CSDs), thereby reducing data movement between storage and system memory. In addition, Smart-Infinity features an efficient data transfer handler to overlap communication with fixed memory usage and employs accelerator-assisted gradient compression and decompression to improve scalability. While these innovations significantly accelerate LLM training, Smart-Infinity relies on specialized hardware, limiting its applicability to systems equipped with NSP-enabled storage. In contrast, MemAscend addresses system memory inefficiencies in general SSD offloading systems like ZeRO-Infinity, which support distributed training from single-node setups to large-scale clusters without dedicated hardware. Thus, Smart-Infinity and MemAscend target different optimization contexts.


\subsubsection{LoHan}
LoHan~\cite{liao2024lohanlowcosthighperformancesystem} is tailored for fine-tuning very large models on a single consumer-grade GPU within a commodity server with limited main memory. It achieves this through holistic tensor management, introducing techniques such as active gradient offloading, which overlaps CPU-based optimizer execution with the GPU’s backward pass, and traffic-aware activation swapping across GPU, CPU, and SSD tiers. LoHan demonstrates impressive capabilities under extreme resource constraints (e.g., fine-tuning a 175B model on a single RTX 4090). However, its design is primarily optimized for single-GPU environments, with only brief consideration for multi-GPU setups. In contrast, MemAscend targets system memory inefficiencies in SSD offloading systems like ZeRO-Infinity, designed for distributed training across a broader range of hardware scales. Therefore, LoHan and MemAscend serve distinct use cases within the LLM training optimization.


To clarify the positioning of MemAscend relative to contemporary SSD offloading techniques, Table~\ref{tab:comparison_related_works} provides a comparative summary.

\vspace{-0.15in}

\begin{table}[h]
\caption{Comparison of MemAscend with recent SSD offloading techniques.}
\label{tab:comparison_related_works}
\centering
\footnotesize
\renewcommand{\arraystretch}{1.1}

\resizebox{0.85\columnwidth}{!}{
\begin{tabularx}{\columnwidth}{@{} l >{\raggedright\arraybackslash}X l @{}}
\toprule
\textbf{Technique} & \textbf{Primary Scope and Memory Efficiency} & \textbf{Hardware} \\
\midrule
ZeRO-Infinity & General distributed offloading, but it suffers from memory fragmentation and overhead spikes. & \makecell[l]{Commodity \\ (Multi-GPU)} \\
\addlinespace
LoHan & Minimizes host memory usage through integrated SSD activation offloading and employs holistic scheduling for optimal GPU utilization in single-GPU scenarios. & \makecell[l]{Commodity \\ (Single-GPU)} \\
\addlinespace
Smart-Infinity & Near-storage processing for optimizer updates, but it does not address host DRAM bottlenecks. & NSP-enabled SSD \\
\addlinespace
\textbf{MemAscend} & \textbf{Optimizes general distributed fine-tuning and resolves ZeRO-Infinity's inefficiencies.} & \makecell[l]{\textbf{Commodity} \\ \textbf{(Multi-GPU)}} \\
\bottomrule
\end{tabularx}
}
\vspace{-0.15in}
\end{table}

\vspace{-0.1in}
\subsection{GPU Memory Efficiency Optimizations}
\label{subsec:gpu_memory_optimizations}

\begin{figure}[t]
    \centering
    \subfloat[Context length=512]{\includegraphics[height=1.6in]{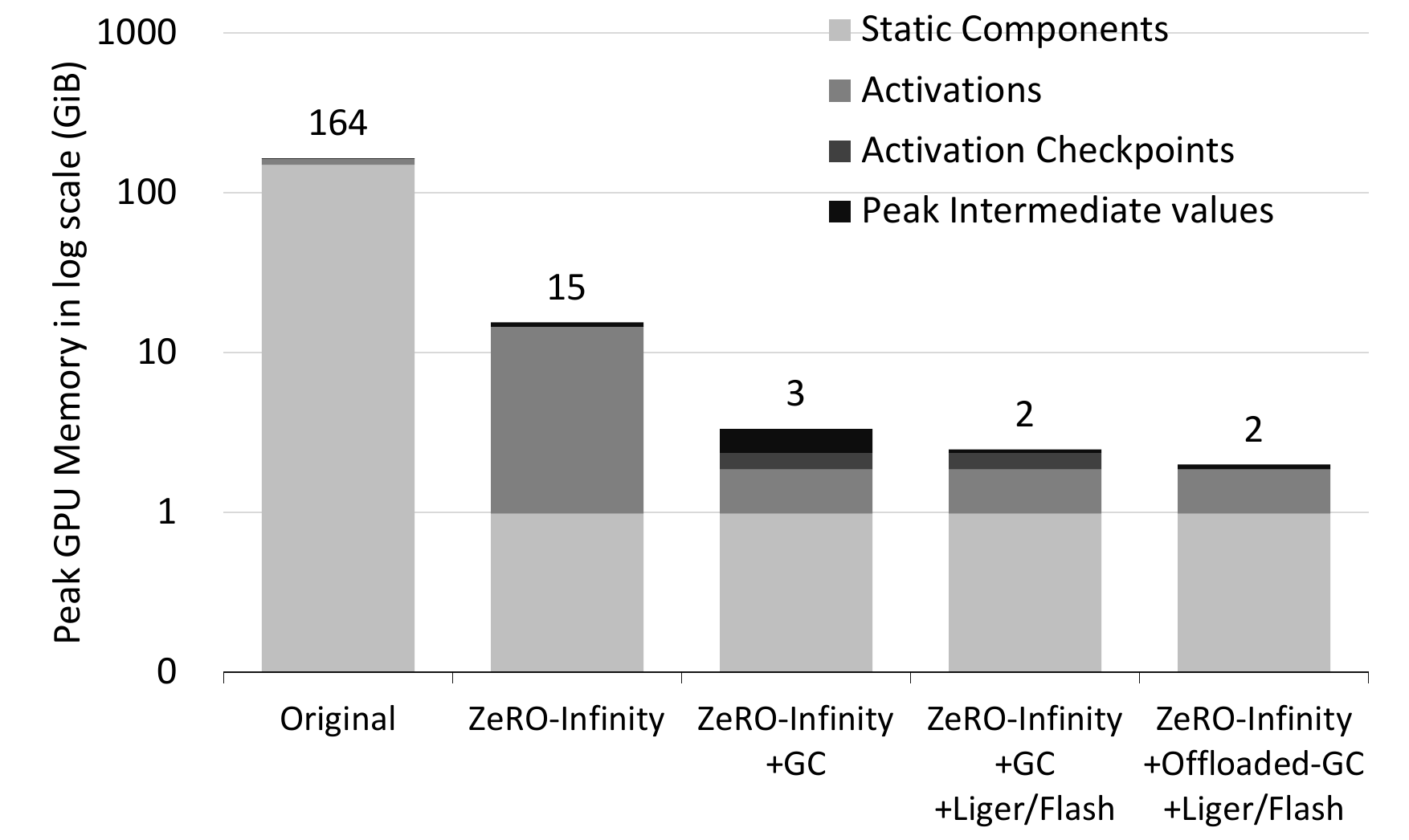}}
    \hfil
    \subfloat[Context length=32768]{\includegraphics[height=1.6in]{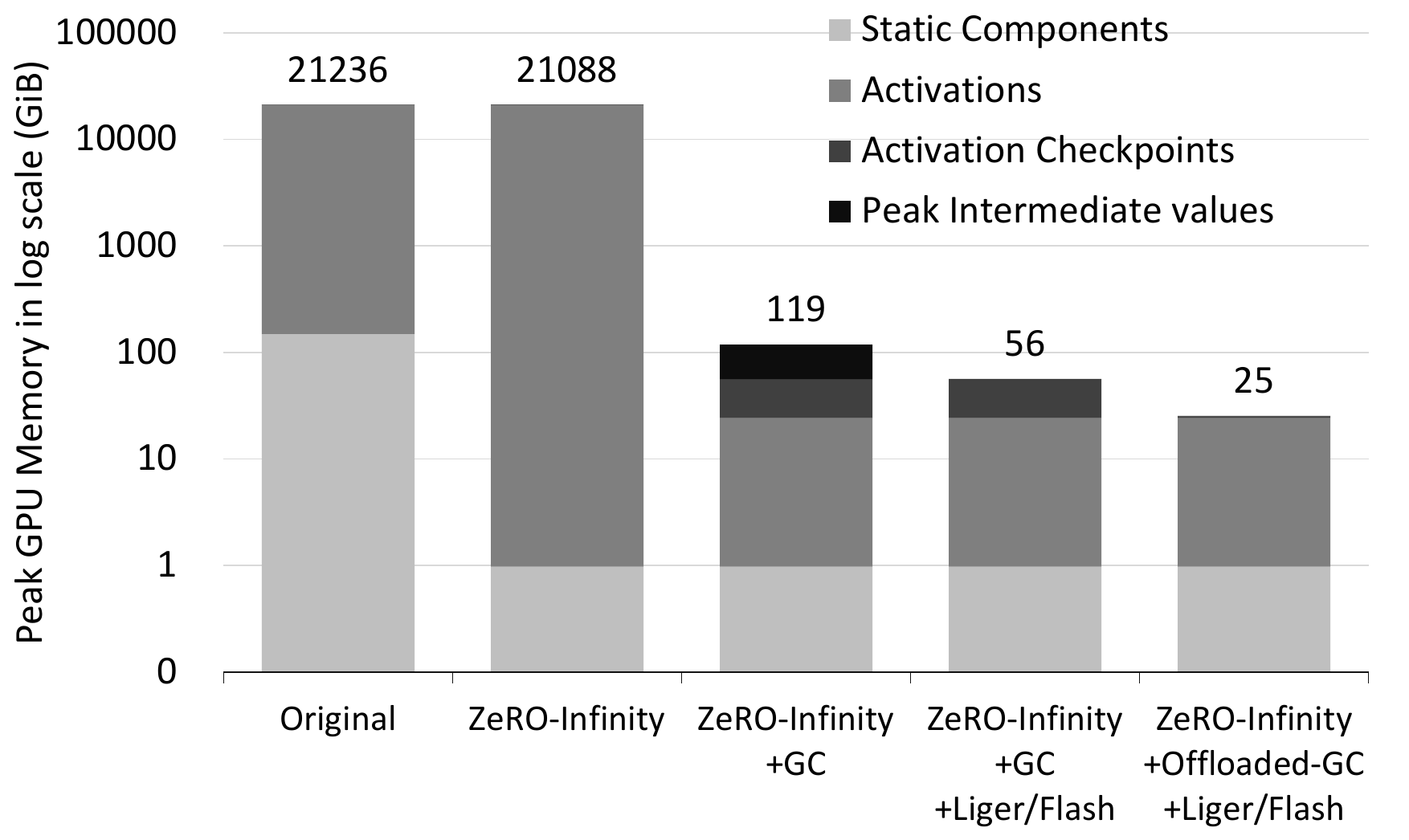}}
    \caption{Comparison of GPU memory usage between short and long context lengths with a batch size of 4 for training an 8-billion-parameter model, across different GPU memory efficiency optimizations. The y-axis uses a base-10 logarithmic scale. GC represents enabled Gradient Checkpointing, Liger/Flash represent enabled Liger-Kernel and FlashAttention, and Offloaded-GC represents enabled Offloaded Gradient Checkpointing.}
    \label{fig:background_vram}
    \vspace{-0.25in}
\end{figure}

Although SSD offloading alleviates the capacity constraints of storing large model states, optimizing the usage of limited, high-bandwidth GPU memory remains essential for maintaining training performance. Several techniques have been proposed to improve GPU memory efficiency, particularly the residual memory, and can be applied alongside SSD offloading systems such as ZeRO-Infinity and MemAscend. By reducing the GPU memory footprint, these methods enable the training of larger models, support increased batch sizes, and allow for longer context lengths, potentially enhancing overall throughput. Otherwise, the SSD offloading still faces the out-of-memory (OOM) issue when increasing batch size or context lengths, since GPU memory cannot accommodate all activations and intermediate values. Detailed GPU memory usage with different memory efficiency methods enabled is shown in Figure~\ref{fig:background_vram}. Accordingly, it can be seen that the GPU memory usage is further reduced after enabling the following optimizations alongside Zero-Infinity. Therefore, \textit{this paper integrates these methods with the state-of-the-art offloading framework (i.e., Zero-Infinity) and utilizes it as the comparison baseline for the proposed MemAscend.}

\subsubsection{Liger-Kernel}
Liger-Kernel~\cite{hsu2025ligerkernelefficienttriton} provides a library of optimized Triton kernels for common operations in LLM training, such as RMSNorm, SwiGLU, and Cross Entropy loss computation. Standard deep learning frameworks like PyTorch typically execute these operations through multiple separate CUDA kernel launches, resulting in additional intermediate memory allocations and kernel launch overhead. Liger-Kernel addresses this by employing kernel fusion, combining multiple operations into a single, optimized kernel. This eliminates the need to materialize large intermediate tensors (e.g., the full logit tensor before softmax in Cross Entropy), thereby reducing peak GPU memory usage and improving execution speed. By integrating Liger-Kernel, GPU computational efficiency is maximized, and memory spikes during forward and backward passes are minimized, making more effective use of GPU resources freed by SSD offloading.


\subsubsection{Flash-Attention}
Flash-Attention~\cite{dao2022flashattentionfastmemoryefficientexact} is an I/O-aware, exact attention algorithm designed to overcome the quadratic time and memory complexity of standard attention mechanisms, especially for long sequences. Instead of materializing the entire $N \times N$ attention matrix in high-bandwidth GPU memory, Flash-Attention computes attention outputs block-by-block using tiling and leverages fast on-chip SRAM. This significantly reduces memory traffic to HBM and lowers memory complexity from $O(N^2)$ to $O(N)$, all while maintaining numerical accuracy without relying on approximation. By dramatically cutting the memory required for attention and accelerating computation, Flash-Attention pairs well with SSD offloading strategies by reducing the GPU memory pressure from dynamic activations.


\subsubsection{Gradient Checkpointing}
Gradient checkpointing~\cite{chen2016trainingdeepnetssublinear}, also known as activation checkpointing or recomputation, reduces GPU memory usage during the forward pass by trading it for additional computation in the backward pass. Instead of storing all intermediate activations needed for backpropagation, only a small subset (checkpoints) is retained. During the backward pass, missing activations are recomputed on-the-fly from the nearest checkpoint. Strategies such as fixed-interval checkpointing (e.g., every $k$ layers) or adaptive checkpointing based on memory cost can be used. This technique significantly reduces peak memory consumption, enabling the training of deeper models or longer sequences.


\subsubsection{Offloading Checkpointed Gradients}
Building on gradient checkpointing, offloading the checkpointed gradient can further reduce GPU memory consumption by transferring the checkpointed activations from GPU memory to CPU system memory. These checkpoints are transferred back to the GPU just in time for recomputation during the backward pass. ZeRO-Infinity includes support for this mechanism, and systems like Unsloth~\cite{unslothai_unsloth_2025} provide efficient implementations using asynchronous transfers. While this technique can greatly expand feasible model sizes and context lengths, it introduces dependency on system memory bandwidth and capacity. Our work, MemAscend, enhances the practicality of offloading checkpointed gradient by reducing system memory fragmentation and overhead within ZeRO-Infinity. The memory reclaimed by MemAscend can be reallocated to store more or larger offloaded checkpoints, thereby supporting larger models or sequences under the same hardware constraints.

In summary, although SSD offloading systems like ZeRO-Infinity lay the groundwork for training massive models using heterogeneous memory, our investigation reveals that substantial inefficiencies remain, particularly in system memory management. The design goal of MemAscend is to directly address these limitations by optimizing system memory usage within SSD offloading systems. Then, through integration with complementary techniques such as SSDTrain, Liger-Kernel, Flash-Attention, and offloaded gradient checkpointing, MemAscend enhances overall memory efficiency and scalability. These improvements allow SSD offloading systems to operate effectively across a wider range of hardware configurations while enabling better synergy with other memory optimization strategies.

\vspace{-0.1in}
\section{Observation and Motivation}\label{sec:observation}
This section presents key observations that reveal critical inefficiencies in existing SSD offloading systems. Despite their ability to extend memory capacity, challenges such as system memory fragmentation, inefficient pinned memory allocation, peak memory bottlenecks, and filesystem-induced I/O overhead remain unresolved. These issues degrade performance and limit scalability. The final subsection outlines the motivation for MemAscend, which aims to address these problems to enhance the efficiency of SSD offloading frameworks.

\vspace{-0.1in}
\subsection{System Memory Fragmentation Issue}
\label{sub:system_memory_fragmentation}

State-of-the-art (SOTA) SSD offloading techniques utilize a parameter buffer pool in system memory to prefetch weights from SSDs during layer execution. However, current designs suffer from memory fragmentation, inflating consumption beyond actual model needs. This occurs because the pool relies on pre-allocated, uniform-sized pinned buffers. When a layer requests weights, it occupies a buffer for SSD-to-Host I/O and subsequent Host-to-Device (H2D) transfer before returning it to the pool. Since prefetching keeps multiple transformer blocks\footnote{Transformer blocks are the fundamental building units of the transformer architecture and are designed to process input sequences by capturing relationships between elements through attention mechanisms and feed-forward networks.} "in flight" simultaneously, the total pool size scales as the product of the buffer size, the buffers required per block, and the number of blocks in flight. Even though GPUDirect Storage (GDS) bypasses system DRAM, the core issue of inefficient buffer management remains. While ZeRO-Infinity uses GDS to load weights directly from NVMe SSDs into GPU memory, it merely relocates the buffer pool from system memory to GPU memory. In resource-constrained fine-tuning, this reserved GPU capacity competes with activations and intermediates, directly limiting batch size, context length, and model scale. Consequently, this work focuses on the system memory setting to address these allocation inefficiencies.

The core issue is that the size of each pinned memory buffer in the buffer pool is fixed based on the largest weight tensor in the model, resulting in severe internal fragmentation. In modern transformer-based models, the embedding layer often contains the largest weights. For example, in LLaMA 3 8B, the embedding layer is sized based on a vocabulary of 128,256 and a hidden dimension of 5,120, making it significantly larger than other layers. In contrast, layers such as the up, gate, and down projections in the feed-forward network (typically sized 14,336 × 5,120) and the key/value projection layers (1,024 × 5,120) are much smaller. This imbalance leads to inefficient memory usage. As a result, although the total buffer pool is sized at 13.05 GiB, the actual memory required to hold the largest amount of tensors in use at any time is only 3.81 GiB, leading to 70.82\% internal fragmentation.

\vspace{-0.1in}
\subsection{Inefficiency in Pinned Memory Allocation}
\label{sec:motivation:pinned_memory}
Pinned (page-locked) system memory underpins high-bandwidth transfers in SSD offloading frameworks. This is because GPUs can only launch PCIe DMA to physical pages that remain resident; all data shuttled between SSD and GPU DRAM must first land in pinned system memory. Pre-allocating these buffers eliminates the extra copy from pageable DRAM to a driver-managed staging area, cutting latency and CPU overhead. Consequently, large tensors that move frequently, model weights fetched from SSD, or gradients written back, are allocated once and reused for the duration of training. The default pinned-memory allocator in PyTorch, \texttt{CachingHostAllocator} rounds every request up to the next power of two to curb fragmentation in highly dynamic workloads. However, that policy backfires in SSD offloading, where allocations are both large and long-lived. Flat buffers for gradient shards, optimizer states, or embedding matrices may be hundreds of MiB to several GiB; aligning a 2.1 GiB request to 4 GiB needlessly wastes almost 2 GiB. Because these buffers persist, the alignment overhead turns into permanent internal fragmentation rather than a transient cost. Eliminating such over-alignment is therefore critical to reclaim system memory and raise the efficiency in SSD offloading systems.

\vspace{-0.1in}
\subsection{Peak Memory Bottlenecks in System Memory}
\label{sec:motivation:peak}
\begin{figure}[h]
    \vspace{-0.1in}
    \centering
    \includegraphics[width=3.3in]{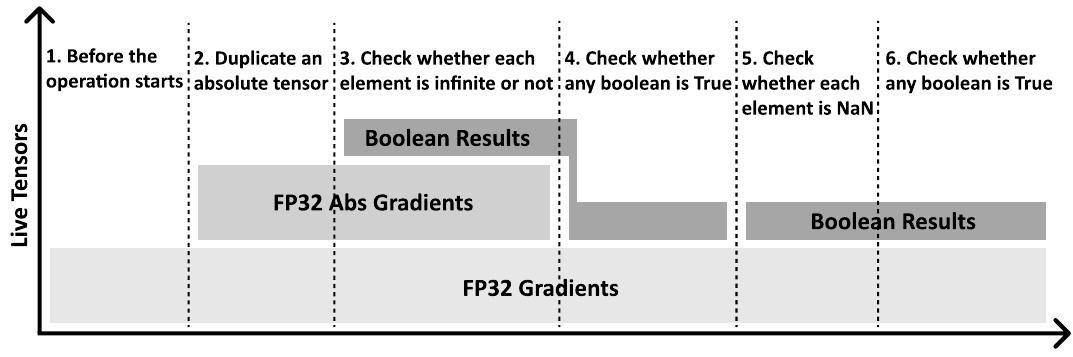}
    \vspace{-0.1in}
    \caption{Tensor lifetimes during gradient overflow checks in ZeRO-Infinity.}\label{fig:fused_overflow_check_method}
    \vspace{-0.1in}
\end{figure}

In ZeRO-Infinity, the flat buffer for gradient partitioning is a single contiguous block that stores all partitioned gradients, enabling streamlined communication and computation during distributed training. When gradient accumulation is enabled, the same buffer is reused across iterations. Because ZeRO-Offload and ZeRO-Infinity place the buffer in system memory, temporary GPU results can be copied into it efficiently. However, a critical drawback is that the buffer’s capacity needs to match the total model-parameter size. To support accumulation, the buffer is allocated in \texttt{fp32} even when forward and backward passes use \texttt{fp16}, so its footprint scales directly with model complexity. As Figure \ref{fig:fused_overflow_check_method} shows, the main peak-memory issue is the gradient-overflow check performed each iteration in mixed-precision training. If any intermediate value overflows, the scaling factor is reduced to avoid instability. This check introduces both memory and latency overhead.

To detect overflows during the overflow check, each value must be examined to determine whether it meets any of the following three conditions: infinity, negative infinity, or NaN. In PyTorch, this involves calling \texttt{isinf()} followed by \texttt{any()}. Internally, \texttt{isinf()} triggers \texttt{isabs()} to duplicate the tensor before performing comparisons, which creates a temporary copy and a Boolean tensor, resulting in a memory spike to about 2.25× the original size, as shown in Steps 2 of Figure~\ref{fig:fused_overflow_check_method}. At Step 3, \texttt{isnan()} avoids \texttt{isabs()} but still produces a Boolean tensor, adding another 1.25× memory peak. For example, when training an 8-billion-parameter model, the flat buffer on the CPU consumes 29.91 GiB, but the overflow check pushes usage to 67.30 GiB. On the other hand, the sequence of operations in the overflow check introduces significant latency: (1) \texttt{isabs()}, (2) \texttt{isinf()}, (3) \texttt{any()}, (4) \texttt{isnan()}, and (5) \texttt{any()}, as shown in Steps 2 to 6 in Figure~\ref{fig:fused_overflow_check_method}. This chain of steps adds roughly 5507 ms per iteration for an 8-billion-parameter model on the Configuration 1 machine described in Table~\ref{tab:hardware_specs}, accounting for 13.36\% of the single iteration latency when training with a context length of 4096 and a batch size of 8, thereby substantially slowing down the training process. Since this check runs every iteration, the cumulative impact on performance is considerable. Thus, both the memory spike and the extra computation make gradient overflow checking a major bottleneck for large-scale models.




\vspace{-0.1in}
\subsection{Filesystem Overhead in SSD offloading}
\label{sec:motivation:filesystem_overhead}
In an SSD offloading system, the SSD serves as extended memory. It stores low-precision compute weights, such as \texttt{fp16}, along with the full-precision \texttt{fp32} master weights, gradients, and optimizer states. These data are loaded or offloaded depending on the computational context. The current design, also known as DeepNVMe in ZeRO Infinity, uses conventional file systems with the \texttt{O\_DIRECT} flag, which bypasses the page cache to ensure that data transfers go directly to the SSD. Each tensor is offloaded to a separate file, allowing file systems, such as ext4, to manage storage allocation without direct intervention. However, writing to ext4 carries overhead. The file system maintains metadata for each file, including name, size, permissions, and timestamps, which requires additional I/O and increases latency. Allocating disk blocks can also involve searching for contiguous space, updating allocation tables, and maintaining alignment, all of which add computational cost and risk fragmentation. When the SSD is used as an extension of memory and subjected to sustained I/O, these file system overheads accumulate. This added latency can limit the efficiency gains expected from offloading, particularly during training.

\vspace{-0.1in}
\subsection{Motivation}
Even though the SSD offloading technique offers a promising solution for training larger models on systems with limited resources, \textit{those observed system-memory bottlenecks intrinsic to SSD offloading lower scalability and erode cost-effectiveness.} To understand the limit of existing SSD offloading techniques, a motivational experiment is conducted based on a machine with 24 GiB of GPU memory and a 128 GiB system memory limit, and results are summarized in Table~\ref{tab:peak_dram_usage}. Accordingly, the All-in-GPU approach, which stores everything on the GPU, supports training models with up to 1 billion parameters. By using ZeRO-Offload, which shifts data to CPU memory, the same system can handle models with up to 3 billion parameters. With the SOTA SSD offloading strategy, ZeRO-Infinity, it can train models with up to 8 billion parameters. It can be observed that, even though the model size is increased, the system memory usage grows more drastically and soon reaches the system limits of commodity CPUs. According to the above observations, inefficiencies such as buffer pool fragmentation, unnecessary alignment for pinned memory allocations, and memory spikes during gradient overflow checks lead to significant waste of system memory and prevent the SOTA SSD offloading strategy from accommodating larger models.

\begin{figure}[t]
    \centering
    \includegraphics[height=1.7in]{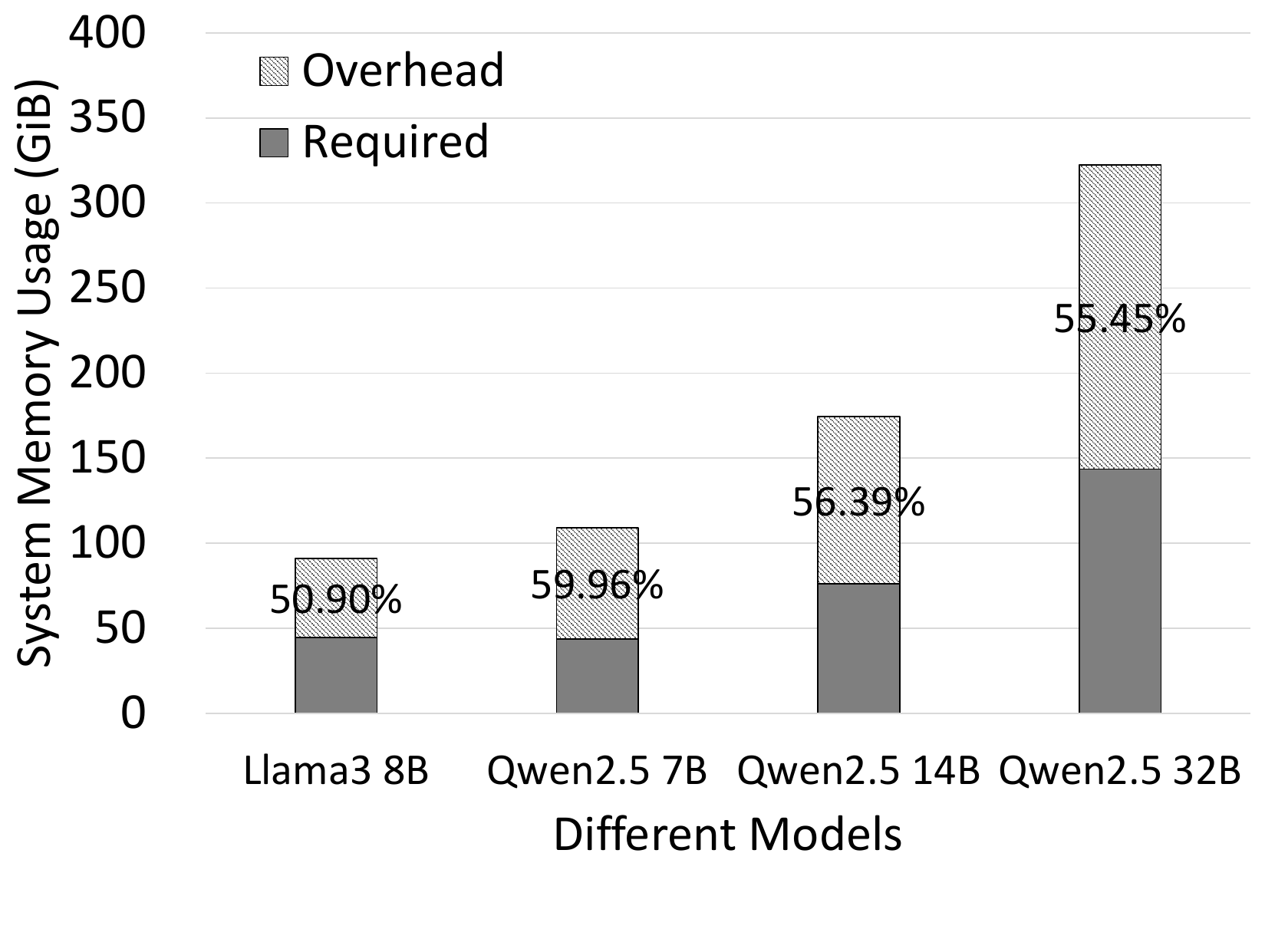}
    \vspace{-0.3in}
    \caption{System memory overhead in the original SSD offloading system across different models. This demonstrates that the original SSD offloading system limited the trainable model size, context length, and batch size due to memory overhead.}
    \label{fig:motivation_memory_waste}
    \vspace{-0.2in}
\end{figure}

To quantify the actual required and wasted system memory usage, another motivational experiments are conducted on Configuration 2, as mentioned in Table~\ref{tab:hardware_specs}. As shown in Figure~\ref{fig:motivation_memory_waste}, the system wastes an average of 55.7\% of its memory across different models, making system memory the next bottleneck. These limitations reduce the scalability and cost-effectiveness of SSD offloading. This study directly addresses those inefficiencies. Reclaiming wasted system memory can either increase the maximum trainable model size on current hardware or reduce the memory required for existing workloads, improving economic efficiency. The recovered capacity can also be used to store additional offloaded activation checkpoints, supporting longer context windows or larger batch sizes and boosting throughput by increasing useful computation per data movement. Finally, minimizing secondary overheads such as I/O latency and overflow check delays further enhances overall efficiency and scalability.

\begin{table}[t]
    \centering
    \caption{Peak system memory usage comparison on different training model size across different training system approaches.}
    \begin{tabular}{|c|c|c|}
        \hline
        \textbf{Type} & \textbf{Model} & \textbf{Peak System Memory (GiB)} \\ \hline
        All in GPU & 1B & 4.48 \\ \hline
        ZeRO-Offload & 1B & 42.99 \\ \hline
        ZeRO-Infinity & 1B & 39.04 \\ \hline
        All in GPU & 3B & N/A (VRAM OOM) \\ \hline
        ZeRO-Offload & 3B & 104.17 \\ \hline
        ZeRO-Infinity & 3B & 62.97 \\ \hline
        All in GPU & 8B & N/A (VRAM OOM) \\ \hline
        ZeRO-Offload & 8B & N/A (DRAM OOM) \\ \hline
        ZeRO-Infinity & 8B & 91.76 \\ \hline
    \end{tabular}
    \vspace{-0.15in}
    \label{tab:peak_dram_usage}
\end{table}

\begin{figure*}[!t]
\centering
\begin{minipage}[t]{0.66\textwidth}
    \centering
    \includegraphics[height=1.4in]{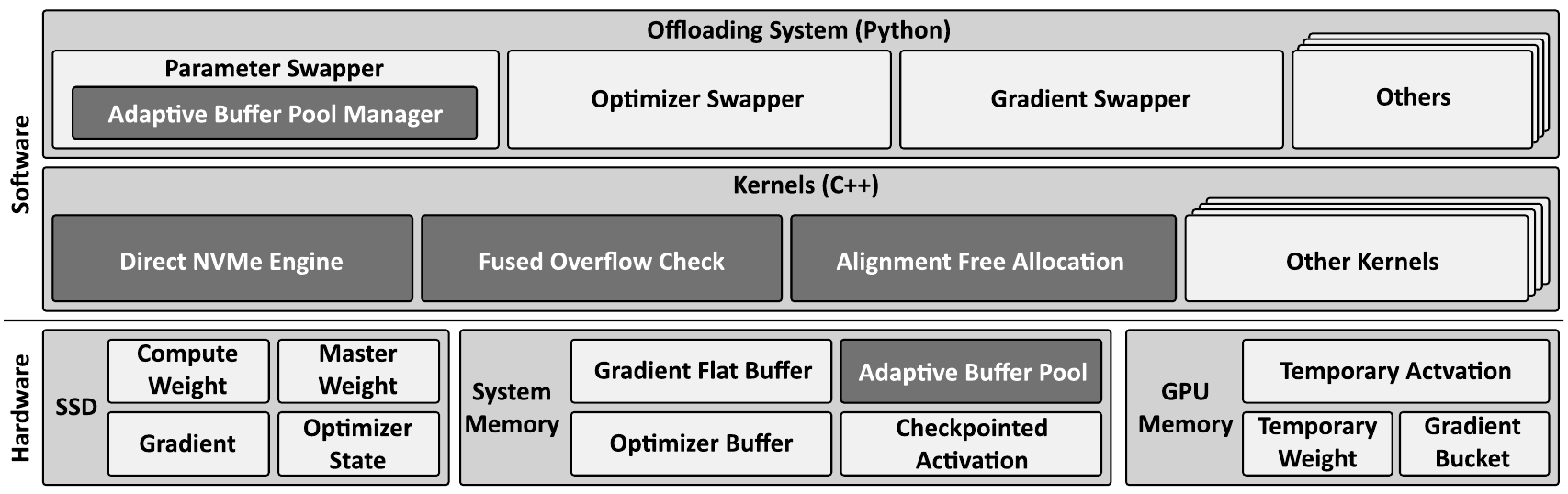}
    \vspace{-0.1in}
    \caption{System architecture of MemAscend, highlighting optimized low-level kernels and memory management for reduced memory usage and enhanced performance.}
    \label{system_architecture}
\end{minipage}
\hfill
\begin{minipage}[t]{0.33\textwidth}
    \vspace{-1.4in}
    \centering
    \subfloat[Monolithic buffer pool in ZeRO-Infinity]{
        \includegraphics[width=0.98\linewidth]{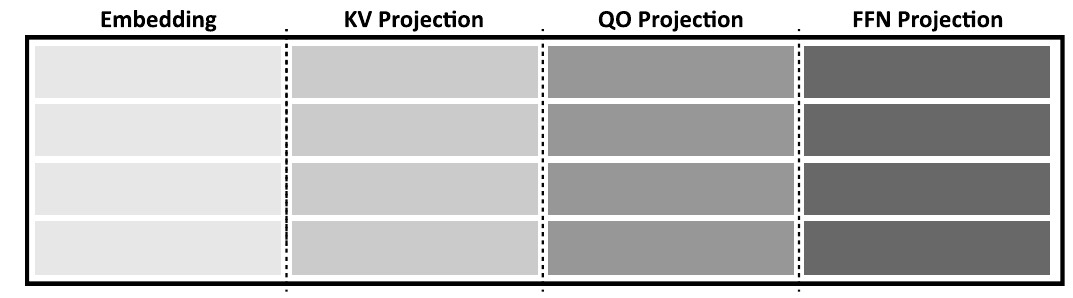}
    }
    \vspace{0.05in}
    \subfloat[Adaptive Buffer Pool in MemAscend]{
        \includegraphics[width=0.98\linewidth]{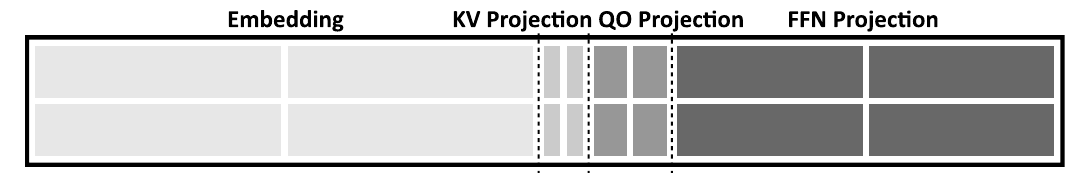}
    }
    \caption{Comparison of parameter buffer pool management in SSD offloading.}
    \label{fig:adaptive_buffer_pool}
\end{minipage}
\vspace{-0.2in}
\end{figure*}

\vspace{-0.1in}
\section{MemAscend}\label{sec:method}


\vspace{-0.1in}
\subsection{Overview}
\label{subsec:overview}

In this study, to resolve the critical issue of system memory inefficiencies in SSD offloading systems, MemAscend is proposed with four key optimizations, which are (1) an adaptive buffer pool, (2) an alignment-free pinned memory allocation, (3) a fused overflow check mechanism, and (4) a direct NVMe engine. Figure~\ref{system_architecture} depicts a software–hardware split, and highlighted components are implemented as part of the proposed MemAscend. The software component represents the primary offloading system written in Python, along with helper C++ kernels that handle low-level operations through a Python-C++ binding interface. This allows the offloading system to perform tasks such as memory allocation and NVMe I/O. The hardware component includes three types of memory devices used in SSD offloading systems: GPU memory, system memory, and SSD. According to the design of MemAscend, three stages of the training process are altered as follows. 

During initialization, gradient-partition buffers, optimizer-state pools, and other critical regions are allocated by the alignment-free pinned memory allocation, eliminating padding waste and fragmentation of PyTorch's default pinned memory mechanism. The forward-and-backward pass then operates in a pipeline: weights are prefetched from SSD to CPU DRAM, moved to the GPU for the all-gather operation. A hook linked to this transfer activates the parameter swapper, which moves tensors from SSD to system memory. The swapper draws exactly-sized chunks from the adaptive buffer pool and hands them to the direct NVMe engine for low-overhead I/O, releasing each chunk once its data reaches GPU memory. The above procedure cycles repeatedly across layers during training. In the final stage, before the optimizer step, a single-pass fused overflow check validates accumulated gradients with negligible extra memory or compute. Notably, the proposed MemAscend is seamlessly integrated with ZeRO-Infinity, Liger-Kernel, FlashAttention, gradient checkpointing, and offloaded gradient checkpointing, ensuring that every byte recovered by MemAscend can be redirected to larger batch sizes or longer context windows, boosting both the throughput and model capacity.

\vspace{-0.1in}
\subsection{Adaptive Buffer Pool}
\label{sec:method:adaptive_buffer_pool}
The system memory requirements of SSD offloading frameworks are primarily driven by data transfer demands between SSDs and GPUs during layer-wise forward and backward propagation, and commonly employ asynchronous prefetching. Due to prefetching operations that will cause \textit{N} number of transformer blocks to be concurrently in flight, the required buffer capacity is therefore the size of \textit{N} transformer blocks. In each transformer block, since smaller tensors can remain permanently in CPU memory, the primary tensors requiring transfer include the feedforward layer weights, specifically the up, down, and gate projections, and the attention layer weights, namely the Q, K, V, and O projections. Additionally, dedicated buffers are required for the embedding layer weights and the final linear layer weights, which are often referred to as the LM head (language-modeling head) and take hidden states from the last transformer block to produce one logit for every token in the vocabulary. Thus, the overall buffering strategy needs to accommodate weights for one embedding layer, \textit{N} transformer blocks each containing weight tensors, and the LM head.

To accommodate the above buffer requirements at the system memory level, a buffer pool design is included in the original SSD offloading architecture. Nonetheless, as shown in Figure~\ref{fig:adaptive_buffer_pool}(a), since the original design fixed the buffer block size to the largest buffer component, the issue of international fragmentation emerges. In this study, as illustrated in Figure~\ref{fig:adaptive_buffer_pool}(b), the adaptive buffer pool design is included to assign separate subpools to tensors of different shapes, avoiding the internal fragmentation that plagues the original monolithic pool in SSD offloading. According to the flow of fine-tuning, four pools are sufficient: one for embedding-related weights, one for feed-forward projections, one for the K and V projections (identical in grouped-query attention), and one for the Q and O projections. Embedding and LM-head weights share the same dimensions—vocabulary × hidden—while feed-forward projections keep the intermediate × hidden shape, so fixed pool dimensions eliminate wasted space. The resulting subgroup counts are $2$, $3N$, $2N$, and $2N$, respectively, which match the exact number of tensors and effectively remove internal fragmentation, with \textit{N} denoting the number of transformer blocks that must be in flight due to prefetching. In this study, the adaptive buffer pool operates entirely on top of PyTorch’s allocator, modifying only logical buffer management. It preserves PyTorch’s safety and compatibility while reducing pinned-memory pressure, scaling effectively to multi-GPU training.

Notably, the management cost of the adaptive buffer pool is kept minimal by following the same underlying allocation approach as ZeRO-Infinity. This approach first allocates a single large, sequential monolithic tensor and then maintains metadata to manage which portion is assigned to each buffer. This design minimizes the allocation overhead that might occur with multi-pool management. When a buffer request occurs, such as when a tensor needs to be fetched from the SSD and copied to the CPU, a hashtable records a unique identification key. This key maps to the data structure containing the buffer's metadata, which includes its size and the index of its portion within the large monolithic tensor. Since the number of distinct buffers is typically small (fewer than 100 in most mainstream open-source models, depending on the configuration), the complexity overhead of managing this hashtable metadata is negligible in terms of both computation and memory usage. Furthermore, this scheme efficiently generalizes to multi-GPU deployments that use parameter partitioning. In such a setup, each process loads only its own parameter shard, causing the per-process buffers to shrink proportionally with the number of partitions. 
Notably, the adaptive buffer pool extends naturally to GDS-based offloading. Since GDS necessitates similar buffer management on the GPU, this method provides a robust framework for minimizing VRAM overhead, demonstrating its broad applicability across different memory tiers.

\vspace{-0.1in}
\subsection{Alignment Free Pinned Memory Allocation}
\label{sec:method:zero_overhead_pinned_memory}

As noted in Section~\ref{sec:motivation:pinned_memory}, PyTorch’s standard caching allocator applies a power-of-two alignment policy. This policy incurs substantial memory waste when managing the large, static tensors essential for SSD offloading. These tensors are allocated once at initialization and persist until training completes, and they are a major contributor to overhead when power-of-two alignment is enforced due to their large size. These critical, parameter-related buffers include: (1) the gradient flat buffer for storing GPU-computed gradients during the backward pass, (2) the parameter buffer pool for prefetching SSD-resident parameters, and (3) the optimizer state buffer for fetching and updating optimizer state subgroups on the CPU. To mitigate this waste, MemAscend uses a custom C++ extension to bypass PyTorch's default alignment.

\textbf{Direct Memory Allocation:} The extension uses \texttt{posix\_memalign} to allocate CPU memory with 4096-byte alignment, satisfying DMA requirements while eliminating power-of-two rounding overhead. \textbf{Page-Locking and CUDA Registration:} Allocated regions are page-locked and registered via \texttt{cudaHostRegister} with the \texttt{cudaHostRegisterPortable} flag, enabling high-throughput DMA transfers across all CUDA contexts. \textbf{Pinned Tensor Creation and Lifecycle Management:} MemAscend wraps the memory into a \texttt{torch::from\_blob} tensor with a custom deleter that leverages PyTorch’s reference counting to safely trigger \texttt{cudaHostUnregister} and free exactly once, bypassing the CPU caching allocator to prevent bookkeeping interference and memory leaks.

The aforementioned direct allocation and registration approach ensures that critical pinned buffers, such as the gradient partition flat buffer and the parameter buffer pool, occupy only the required space plus minimal alignment. Notably, this custom method is applied to static, large offloading buffers. Arbitrary, short-lived, or frequently allocated tensors still rely on PyTorch’s standard caching allocator to preserve the original memory management design choice. This strategy allows MemAscend to significantly reduce static overhead without introducing fragmentation risks that could compromise the system's dynamic scalability. Furthermore, since changing the allocation size does not add any extra operations to the memory management workflow, the alignment-free pinned memory allocation introduces no additional overhead compared to the standard caching allocator.

\vspace{-0.1in}
\subsection{Fused Overflow Check Mechanism}
\label{sec:method:fused_overflow_check}
According to the discussion in Section~\ref{sec:motivation:peak}, the SSD offloading system detects gradient overflows by invoking operations, such as \texttt{isnan()} and \texttt{isinf()}, and leads to peak memory usage when transitioning between these functions. Combining these checks into a single pass removes the intermediate tensors each call would otherwise create, lowering peak memory demand. Gradients are produced in \texttt{fp16} during back-propagation, yet the partition’s flat buffer is retained in \texttt{fp32} for accumulation, so \texttt{fp16}-level overflows must still be vetted against \texttt{fp32} representations. Algorithm \ref{alg:fused_overflow_check_method} illustrates the fused procedure: under IEEE 754, overflow manifests when all exponent bits equal 1; NaN values (Not a Number) add a non-zero mantissa, whereas positive or negative infinity has a zero mantissa. Line 2 defines this bit pattern, and lines 5–8 apply bitwise logic to flag any value with an all-ones exponent as overflowed. Each element is independent, enabling wide parallel execution; practical tests show roughly 97\% efficiency for models spanning 1 B–32 B parameters. By preventing intermediate allocations and exploiting bit-level checks, the fused mechanism removes the overflow-detection bottleneck at scale. The proposed fused overflow check mechanism also generalizes efficiently to multi-GPU deployments, where each GPU process manages its own partition and performs the fused check without synchronization. The fused overflow check uses an OpenMP (Open Multi-Processing)-based parallel implementation, which introduces minor runtime overhead due to thread management, synchronization, and scheduling. However, these costs are negligible compared to the overall performance improvement, as the acceleration achieved effectively amortizes the OpenMP overhead.

\begin{algorithm}[h]
    \caption{Fused Overflow Check}
    \footnotesize
    \label{alg:fused_overflow_check_method}
    \begin{algorithmic}[1]
    \Require Gradient tensor $G[0\,\dots\,N-1]$
    \Ensure Boolean \texttt{overflow}
    \State \texttt{overflow} $\gets$ \texttt{False}
    \State \texttt{EXP\_ALL\_ONES\_MASK} $\gets$ \texttt{0x7F800000}
    \For{\textbf{each} $i = 0$ \textbf{to} $N-1$ \textbf{in parallel}}
      \State $bits \gets \Call{reinterpret\_uint32}{G[i]}$
      \If{$(bits \,\&\, \texttt{EXP\_MASK}) = \texttt{EXP\_MASK}$}
        \State \texttt{overflow} $\gets$ \texttt{True}
        \State \textbf{break}  \Comment{Early exit from all threads}
      \EndIf
    \EndFor
    \State \Return \texttt{overflow}
    \end{algorithmic}
\end{algorithm}

\begin{figure}[b]
    \centering
    \vspace{-0.2in}
    \includegraphics[height=2in]{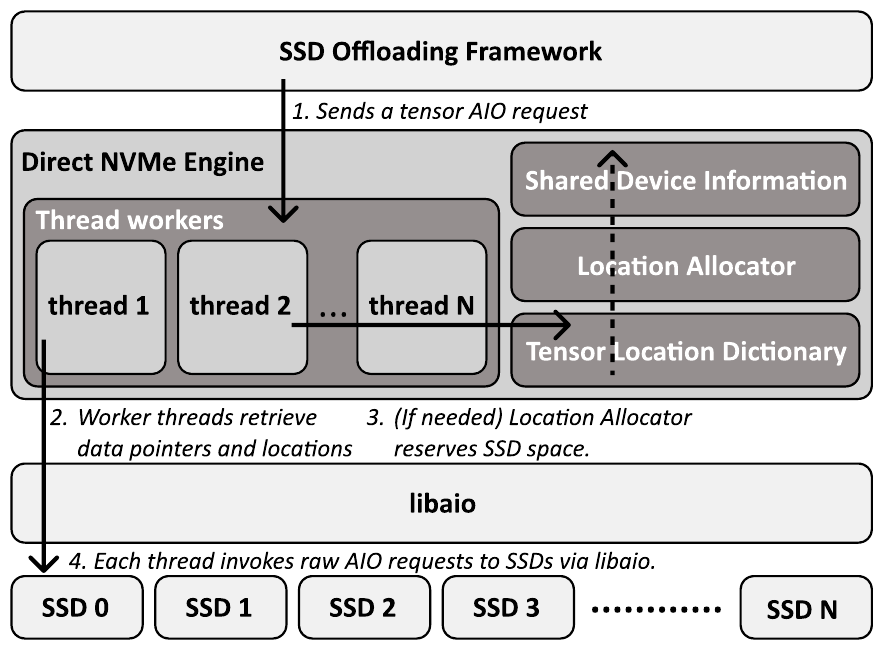}
    \vspace{-0.1in}
    \caption{Workflow of the direct NVMe engine for optimized SSD I/O.}
    \label{direct_nvme_engine}
\end{figure}

\begin{figure*}[!t]
    \centering
    \begin{minipage}{0.32\textwidth}
        \centering
        \includegraphics[height=1.7in]{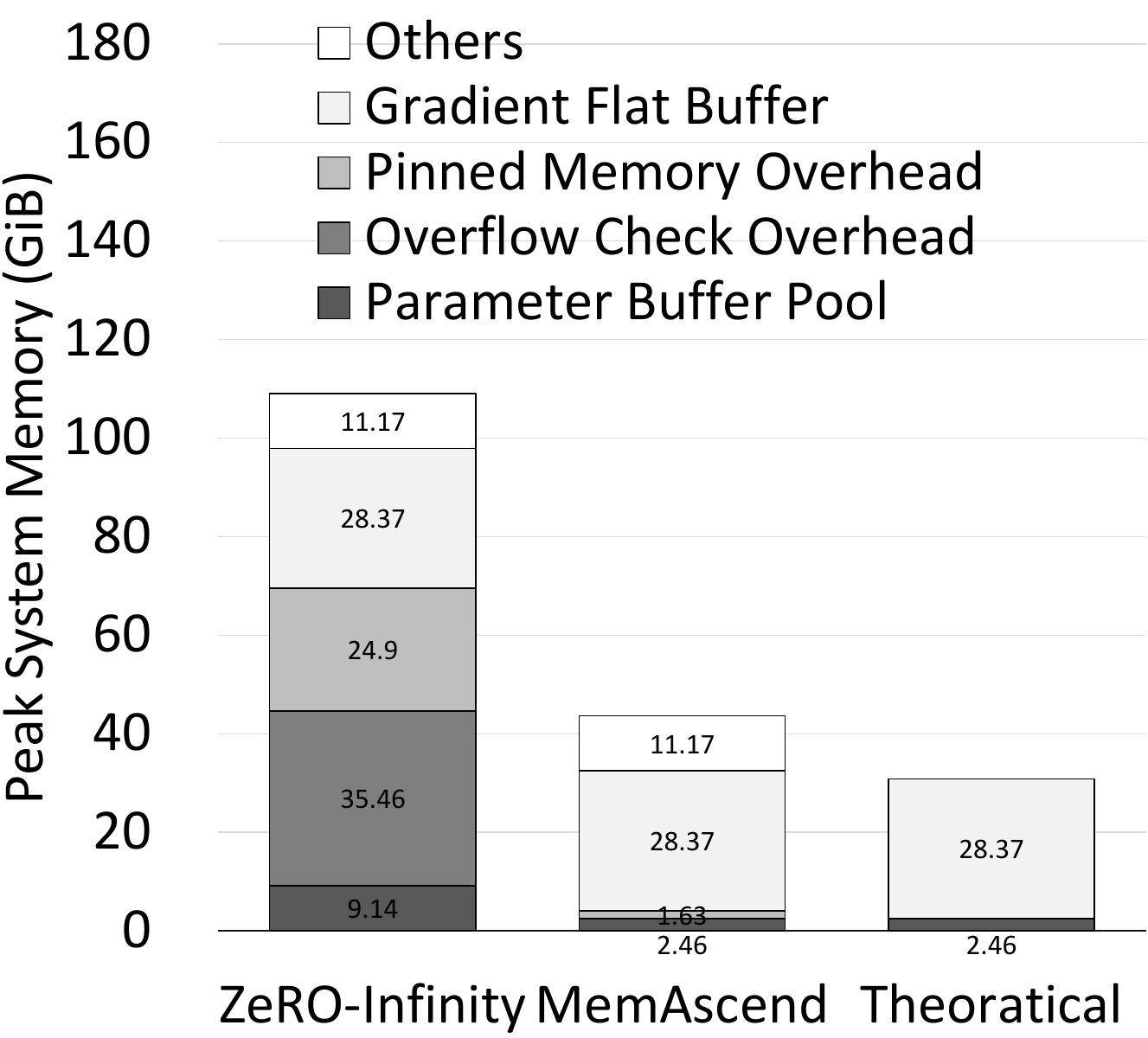}
        \vspace{-0.15in}
        \caption{Peak system memory usage comparison for training Qwen2.5-7B between ZeRO-Infinity, MemAscend and theoretical minimal requirements.}
        \label{fig:analysis_mem_bar}
        \vspace{-0.2in}
    \end{minipage}%
    \hfill
    \begin{minipage}{0.32\textwidth}
        \centering
        \includegraphics[height=1.7in]{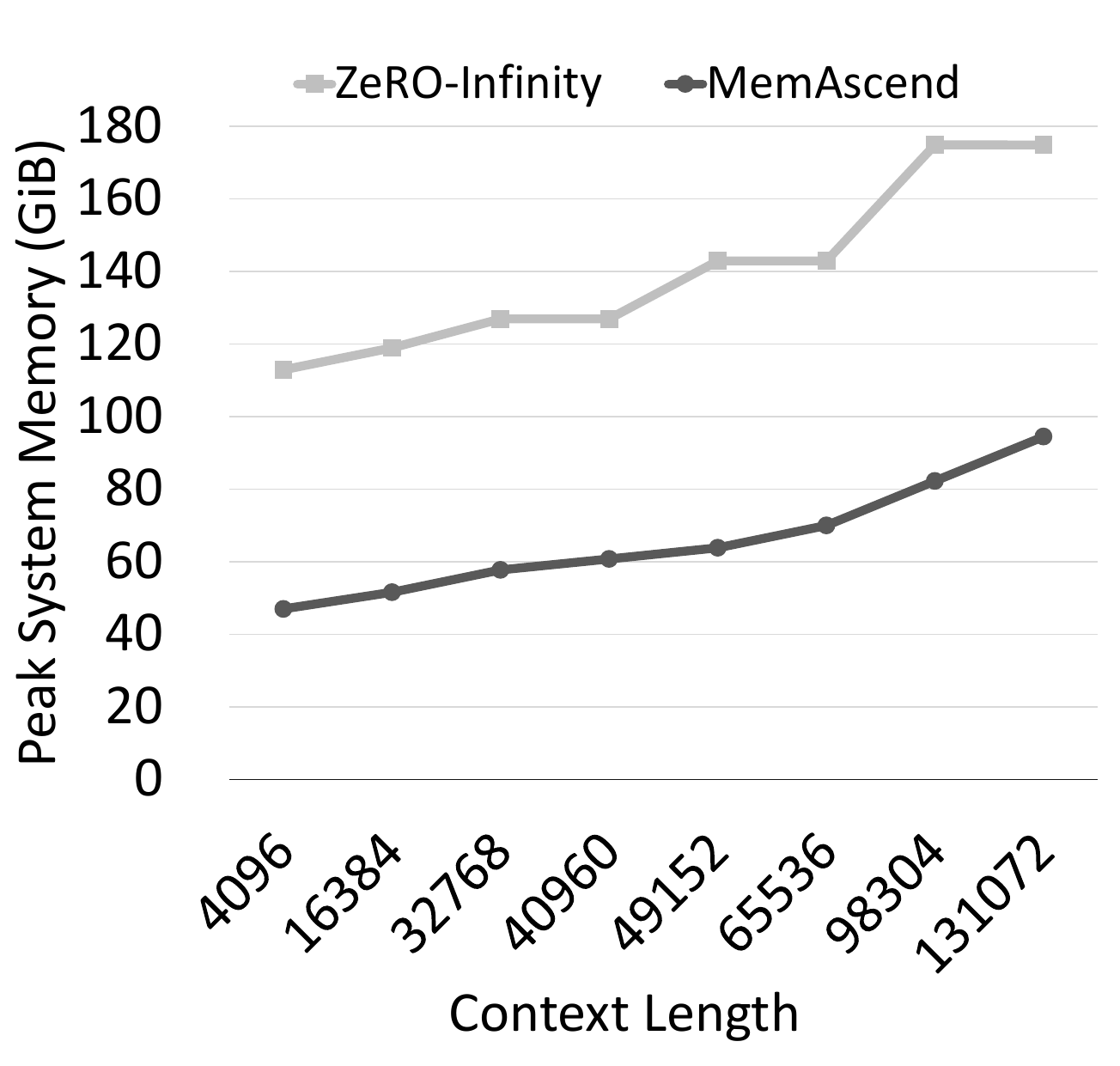}
        \vspace{-0.15in}
        \caption{Peak system memory usage comparison between ZeRO-Infinity and MemAscend across different context lengths.}
        \label{fig:peak_memory_comparison}
        \vspace{-0.2in}
    \end{minipage}%
    \hfill
    \begin{minipage}{0.32\textwidth}
        \vspace{-0.05in}
        \centering
        \includegraphics[height=1.7in]{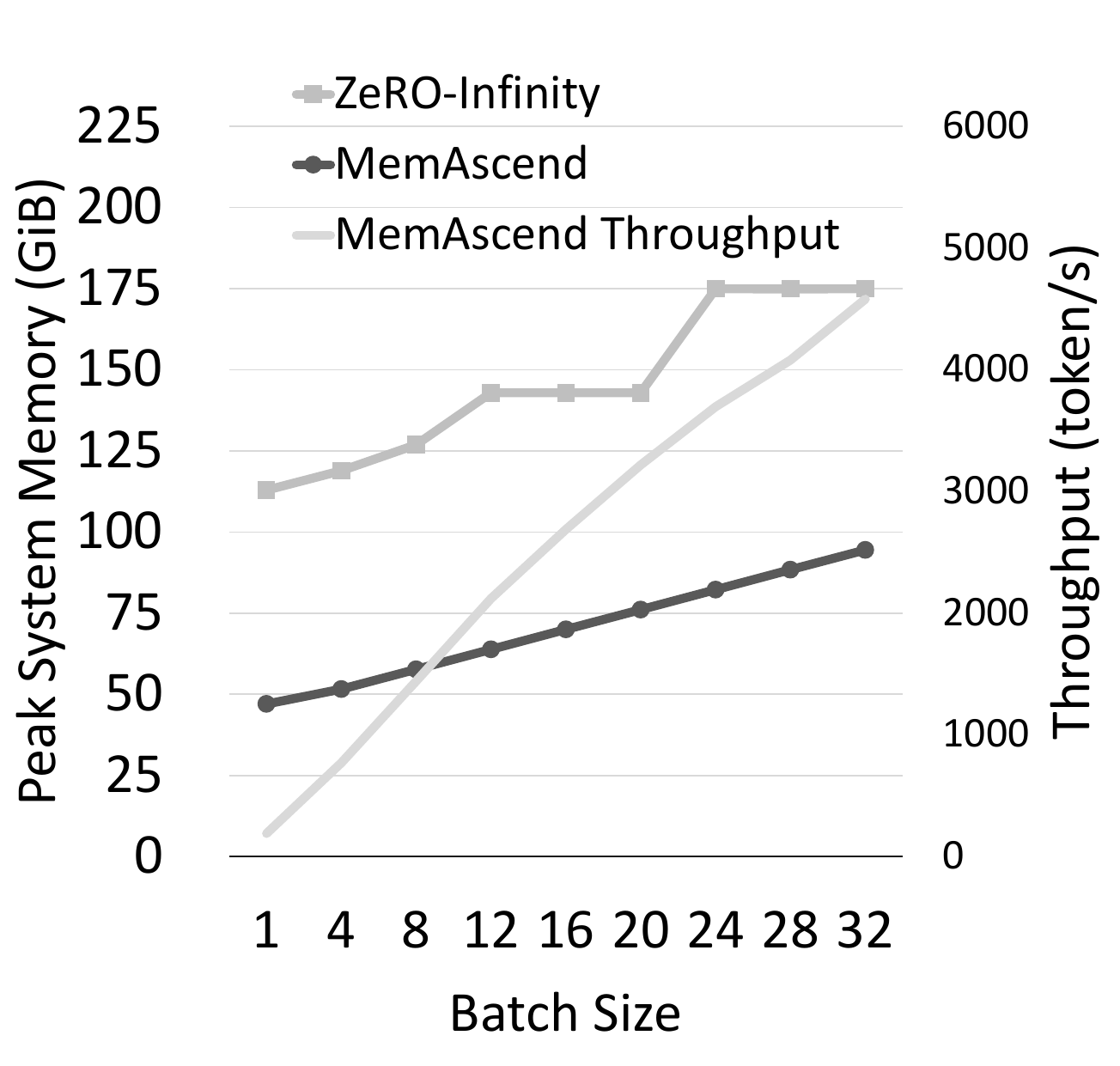}
        \vspace{-0.15in}
        \caption{Peak system memory usage and throughput comparison between ZeRO-Infinity and MemAscend across different batch sizes.}
        \label{fig:throughput_comparison}
        \vspace{-0.2in}
    \end{minipage}
\end{figure*}

\vspace{-0.3in}
\subsection{Direct NVMe Engine}
\label{sec:method:direct_nvme_engine}

The primary goal of the direct NVMe engine is to bypass the Linux filesystem overhead by using raw I/O to directly send asynchronous I/O (AIO) requests to the device driver via \texttt{iouring} or \texttt{libaio}. The overview of architecture and its workflow is shown in Figure~\ref{direct_nvme_engine}. First, the SSD offloading system issues a tensor AIO request. The request context contains a unique tensor name (as a key), a data pointer, and an offset aligned to the tensor size. Next, the direct NVMe engine distributes this request among multiple thread workers, dividing the data into equal portions for balanced performance. Each thread consults the tensor location dictionary to determine which SSD device and logical block address (LBA) to use. If the tensor location dictionary has not recorded the tensor yet, it consults the location allocator for SSD space. This step ensures that the data are horizontally partitioned across SSDs, effectively implementing striping in place of Software RAID 0. The shared device information structure, implemented via shared memory, guarantees that multiple processes do not allocate overlapping SSD regions. Although this requires the overhead of accessing a shared offset counter for the initial allocation of a new, unrecorded tensor, the cost is negligible because it is a simple shared memory integer operation that occurs only once per tensor. After each thread obtains its data pointer, offset, LBA, and raw SSD device path, it sends a raw AIO request through \texttt{libaio}. By applying load balancing, stripping, and multi-threading, the Direct NVMe Engine reduces latency and achieves higher throughput compared to approaches that rely on filesystem-based AIO.

\vspace{-0.1in}
\section{Memory Efficiency and Performance Analysis}
\label{sec:analysis}
By streamlining memory consumption, the MemAscend framework reduces system memory demand relative to earlier SSD offloading techniques. Those lower system memory footprints can translate directly into cheaper training setups; in other words, the same hardware budget can now accommodate models with more trainable parameters. The lower system memory usage also unlocks longer context windows, because the larger activation-checkpoint tensors required for extended sequences can be moved from GPUs to system memory without exceeding capacity. Therefore, instead of slowing down the training process, the overall performance benefits follow as well. Meanwhile, the extra system memory headroom permits larger batch sizes, driving GPUs toward higher utilization and boosting overall throughput. The upcoming analysis examines these gains in detail and explains how the reclaimed RAM supports longer context lengths and bigger batches, both of which contribute to faster training at lower cost. These experiments are conducted using Configuration 1, as described in Table~\ref{tab:hardware_specs}.


\vspace{-0.1in}
\subsection{Memory Efficiency}
\label{sec:analysis:memory_efficiency}
Figure~\ref{fig:analysis_mem_bar} first breaks down system memory usage of the proposed MemAscend while training the \texttt{Qwen2.5-7B} model~\cite{qwen2025qwen25technicalreport}, highlighting both the individual contributions and the synergistic effects of its four optimization components. Notably, both the compared Zero-Infinity and the proposed MemAscend are integrated with the Liger-Kernel, Flash-Attention, and offloaded gradient checkpointing for reducing residual memory usage. First, the adaptive buffer pool reduces the size of the parameter buffer pool by alleviating fragmentation issues and peak buffer usage (for example, from 9.14~GiB to 2.46~GiB, a 73\% reduction), thereby lowering the baseline memory that must be pinned and streamed from the SSD. Second, the alignment-free pinned memory allocation reduces pinned memory overhead (from 24.90~GiB to 1.63~GiB), further decreasing memory usage. This not only resolves the overhead from large components like the gradient flat buffer but also improves the pinned memory overhead caused by the adaptive buffer pool, since it also uses pinned memory for allocation. Third, the direct NVMe engine eliminates filesystem serialization, improving performance for all SSD I/O, including the adaptive buffer pool buffers. Finally, the fused overflow check mechanism removes intermediate buffers, avoiding 35.46~GiB of overflow check overhead. Notably, since each component alleviates bottlenecks that would otherwise reduce the effectiveness of the others, the four optimizations together achieve significantly greater memory efficiency and performance improvements than any one component alone.


In Figure~\ref{fig:analysis_mem_bar}, it can be observed that the power-of-two alignment policy in conventional pinned memory allocation can nearly double memory usage, resulting in 24.90 GiB of allocator-induced overhead. By contrast, MemAscend’s alignment-free pinned memory allocation matches the requested tensor size precisely, lowering this overhead to just 1.63 GiB and achieving a 93.5\% reduction in unnecessary memory consumption. Meanwhile, some components remain constant across methods, including the gradient flat buffer and small system allocations like optimizer state buffers and the swap-out buffer. These account for stable memory usage of roughly 28.37 GiB and 11.17 GiB, respectively. 

On the other hand, to understand the theoretical minimal system memory that would still sustain equivalent throughput, Figure~\ref{fig:analysis_mem_bar} presents the lower-bound memory requirement for ZeRO-Infinity and MemAscend. In this theoretical case, only the parameter buffer pool and the gradient flat buffer are strictly necessary: the former for streaming weights during training and the latter to avoid excessive I/O that would harm throughput. All framework-induced overheads—pinned memory, fragmentation, auxiliary allocations—are excluded. Based on this baseline, MemAscend lowers peak memory from 109.04 GiB (ZeRO-Infinity) to 43.64 GiB (60\% reduction), significantly narrowing the gap to the theoretical minimum. The remaining margin for MemAscend is 12.81 GiB (29\%), while ZeRO-Infinity still requires a 72\% reduction to match this bound. This reduction improves scalability to larger models and longer sequence lengths under fixed memory budgets.

\vspace{-0.1in}
\subsection{Enabling Larger Context Length}
\label{sec:analysis:enable_longer_context}
In the SSD offloading scenario, the key constraint in expanding the context length or batch size is the activation checkpoint value, which scales linearly with both parameters. Within the SSD offloading framework, to alleviate this GPU memory bottleneck, a pinned memory buffer on the CPU can be utilized for swapping activation checkpoint values from GPU memory to system memory, enabling longer context lengths. However, this shifts the bottleneck from GPU memory to system memory. Notably, the total size of activation checkpointed values in system memory is calculated using the following formula, where \(N_g\) denotes the GPU count, \(B\) the batch size, \(C\) the context length, \(L\) the number of layers, \(H\) the hidden size, \(F_{16}\) as the bytes per \texttt{fp16}, and \(P_m\) the pinned memory overhead. Here, the number of layers \(L\) accounts for the activation checkpointing enabled in each transformer layer.

\begin{equation}
\label{eq:activation_checkpoint}
    N_g \times B \times C \times L \times H \times F_{16} + P_m
\end{equation}

Figure~\ref{fig:peak_memory_comparison} shows that the usage of system memory increases with longer context length when running \texttt{Qwen2.5-7B} on two GPUs. Our MemAscend method scales linearly with context length, exhibiting minimal pinned memory overhead, meaning the increase is almost entirely due to context length scaling. In contrast, the original SSD offloading system scales more rapidly and begins with higher memory usage. This disparity arises because its pinned overhead aligns to the nearest power-of-two value, resulting in faster scaling and instances where different context lengths yield identical memory usage due to this alignment. Consequently, our method significantly reduces system memory consumption and slows the scaling rate with increasing context length, offering greater flexibility to extend context length without exhausting system memory. For example, with a 128 GiB system memory limit, the original SSD offloading system supports a context length of 16,384, while MemAscend scales up to 131,072 under the same configuration.

\vspace{-0.1in}
\subsection{Enabling Higher Throughput}
\label{sec:analysis:enable_large_batch_size}
The proposed MemAscend not only supports a larger trainable context length but also unlocks the potential for higher throughput (tokens/sec) by accommodating larger batch sizes through reclaimed system memory budget. Larger batch sizes allow GPUs to process more tokens under the same amount of model weights transfers, thus improving the compute-to-transfer ratio and offsetting SSD offloading overhead. Figure \ref{fig:throughput_comparison} compares the baseline SSD offloading technique with MemAscend during \texttt{Qwen2.5-7B} training on two GPUs. Memory-usage trends remain proportional to batch size or context length, yet MemAscend’s curve rises far more gently, leaving headroom to keep increasing the batch without exhausting host RAM. Consequently, the throughput of MemAscend scales almost linearly with batch size, confirming that more samples per offload cycle translate into higher overall performance. For instance, under a 128 GiB system-memory cap, the baseline tops out at batch 4, whereas MemAscend reaches 32, converting the reclaimed RAM into an eight-fold boost in tokens per second.

\vspace{-0.1in}
\section{Experimental Evaluation}
\label{sec:evaluation}
\subsection{Setup}

The detailed hardware specifications for running the experiments are provided in Table~\ref{tab:hardware_specs}. MemAscend is evaluated using mixed-precision training of large language models with SSD offloading. The workloads vary by model architecture, context length, and batch size. The models include \texttt{Llama3.1-8B} and \texttt{Qwen2.5}, with parameter sizes of 7 billion, 14 billion, and 32 billion for most of our experiments. To test the diversity of model architectures, the model \texttt{Qwen3-30B-A3B} is also included to demonstrate a modern sparse language model. All models use the Hugging Face Transformers library (v4.51.0). The context lengths range from 4,096 to 131,072 tokens. Batch sizes vary across different experiments. The main evaluation metric is peak system memory usage during training and training throughput in tokens per second to assess the overall performance. 

\begin{figure*}[!t]
    \centering
    \begin{minipage}{0.23\textwidth}
        \vspace{-0.1in}
        \centering
        \includegraphics[height=1.6in]{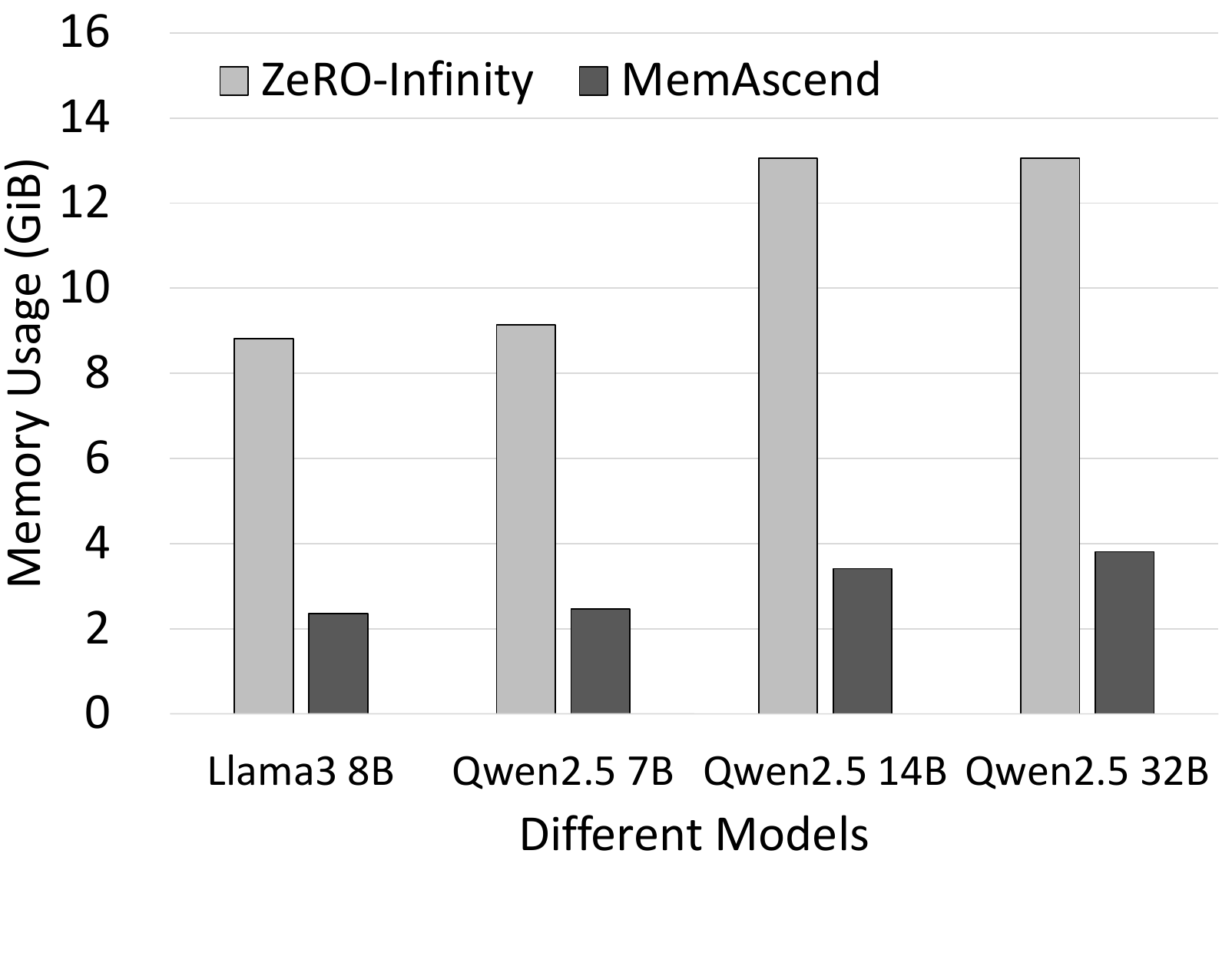}
        \vspace{-0.15in}
        \caption{Parameter buffer pool system memory usage comparison between ZeRO-Infinity and MemAscend across different LLM models.}
        \label{fig:evaluations_adaptive-buffer-pool}
        \vspace{-0.3in}
    \end{minipage}%
    \hfill
    \begin{minipage}{0.51\textwidth}
        \vspace{-0.25in}
        \centering
        \subfloat[Configuration 1 with an Intel(R) Xeon(R) 6780E CPU]{\includegraphics[height=1.6in]{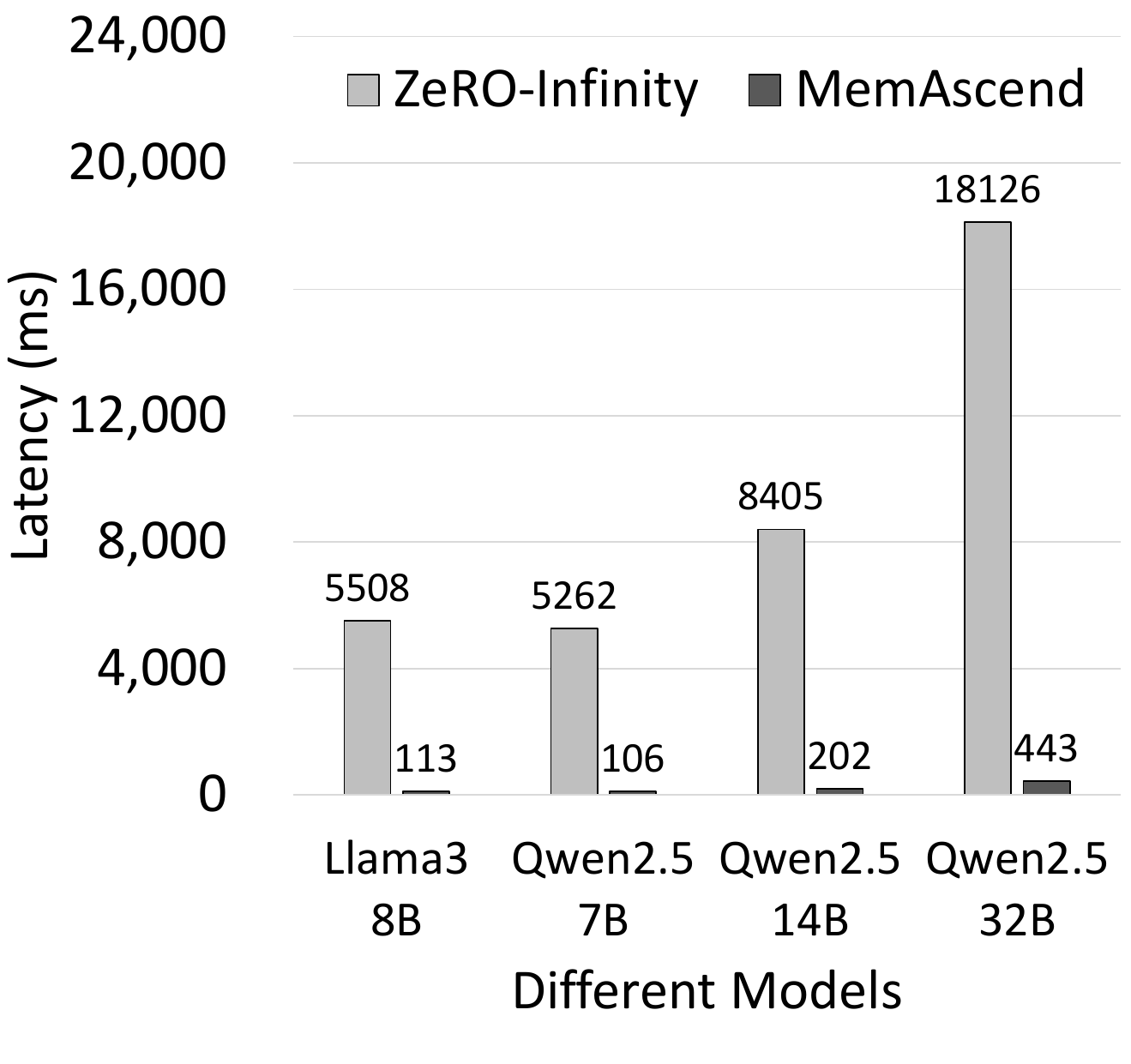}}
        \label{fig_first_case}
        \vspace{-0.03in}
        \hfil
        \subfloat[Configuration 2 with dual AMD EPYC 7282 CPUs]{\includegraphics[height=1.6in]{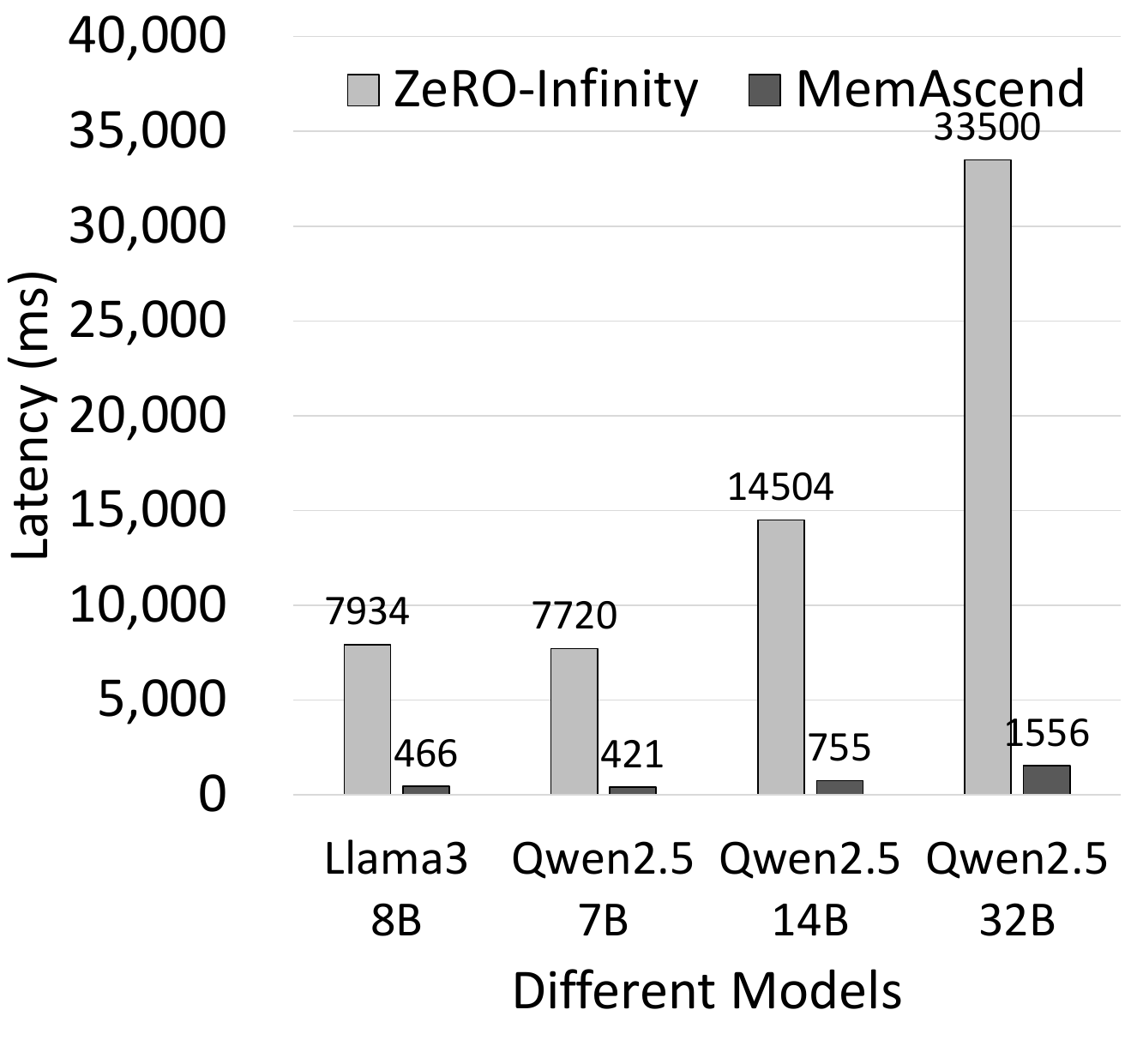}}
        \label{fig_second_case}
        \vspace{-0.03in}
        \caption{Comparison of overflow check latency between ZeRO-Infinity and MemAscend across different CPU types and various models.}
        \vspace{-0.3in}
        \label{fig:fused_overflow_check}
    \end{minipage}%
    \hfill
    \begin{minipage}{0.23\textwidth}
        \vspace{-0.1in}
        \centering
        \includegraphics[height=1.6in]{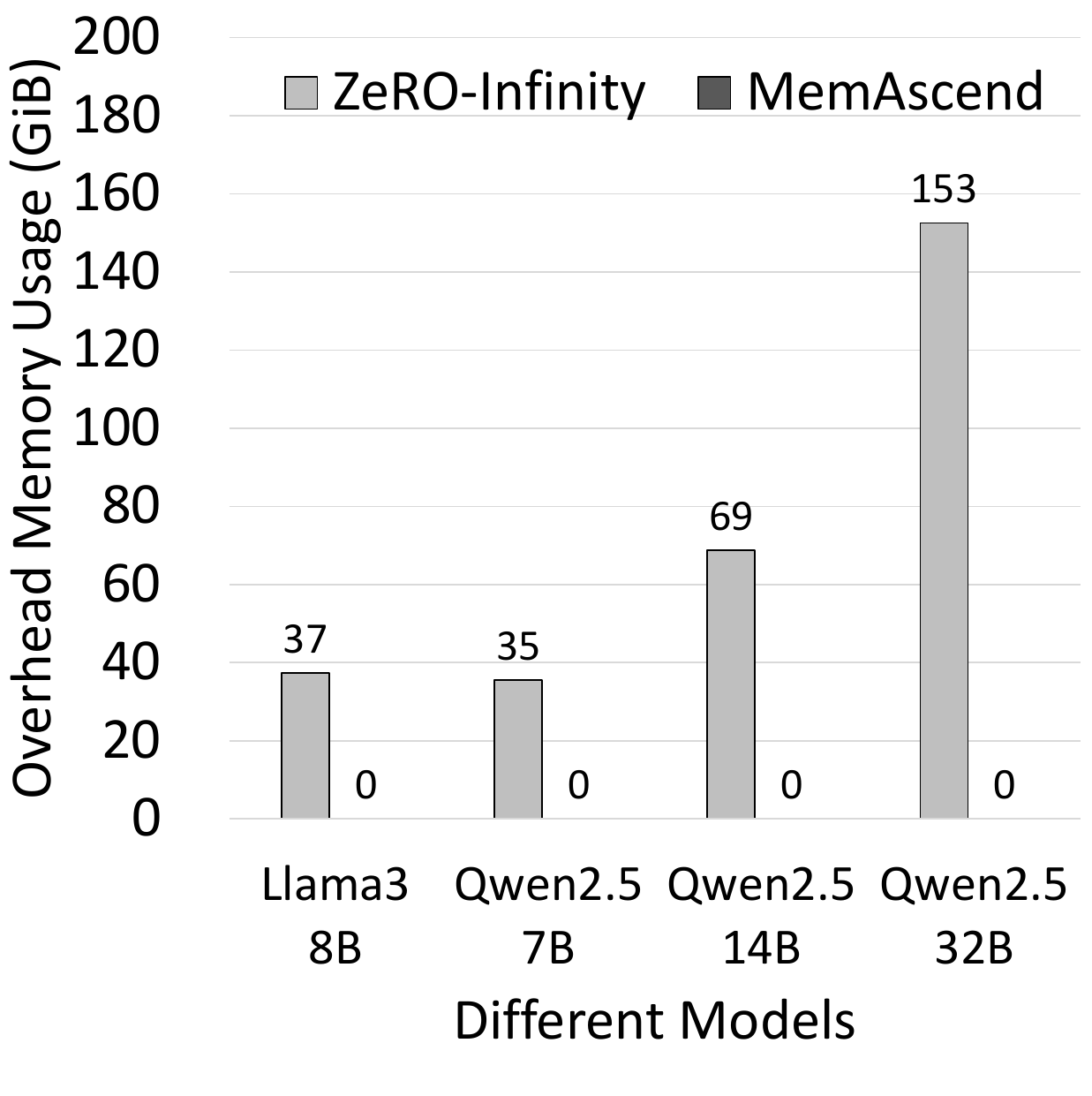}
        \vspace{-0.15in}
        \caption{Comparison of memory overhead for overflow checks between ZeRO-Infinity and MemAscend across various models.}
        \label{fig:fused_overflow_check_memory_overhead}
        \vspace{-0.3in}
    \end{minipage}
    \vspace{0.1in}
\end{figure*}

\begin{table}[h]
\vspace{-0.1in}
\caption{Two Hardware Specifications}
\label{tab:hardware_specs}
\centering
\resizebox{\columnwidth}{!}{
\begin{tabular}{|c||c|c|}
    \hline
    \textbf{Component} & \textbf{Configuration 1} & \textbf{Configuration 2} \\
    \hline
    CPU & Intel Xeon 6780E & 2 $\times$ AMD EPYC 7282 \\
    \hline
    CPU Memory & 1 TB 6400 MT/s DDR5 & 1 TB 3200 MT/s DDR4 \\
    \hline
    PCIe & Gen5 & Gen4 \\
    \hline
    GPU & 2 $\times$ NVIDIA H100 PCIe (w/o NVL) & 1 $\times$ NVIDIA A5000 \\
    \hline
    SSD & 1 $\times$ DapuStor Haishen5 H5100 (7.5 TB) & 2 $\times$ Phison AI100E \\
    \hline
\end{tabular}
}
\vspace{-0.1in}
\end{table}

To evaluate the effectiveness of MemAscend, the SOTA ZeRO-Infinity is compared, and both systems share a common setup to ensure fairness. Static memory components, including model parameters and optimizer states, are stored entirely on the SSD. To reduce GPU memory usage, gradient checkpointing, Flash-Attention, and the Liger-Kernel are enabled for both setups. For experiments that require offloaded gradient checkpointing, a modified version of Unsloth is utilized, adapted from its original single-GPU design in which the original Unsloth implementation of offloaded activation checkpointing only supported a single-GPU setup and hardcoded the CUDA device index in its source code. The modification replaces the hardcoded index with the correct device index for each process in a multi-GPU scenario.

\vspace{-0.15in}
\subsection{Experimental Results}

\subsubsection{Assessment of Individual Methods}
\paragraph{Adaptive Buffer Pool}
As each model varies in weight tensor size due to differences in vocabulary size, hidden size, and other factors, this experimental section presents the memory usage of the buffer pool across various models. The aim is to illustrate the effectiveness of the proposed adaptive buffer pool in reducing memory usage in different situations. The comparison is shown in Figure~\ref{fig:evaluations_adaptive-buffer-pool}. In these configurations, the adaptive buffer pool consistently reduces memory usage by adjusting to the specific tensor sizes of each model, achieving an average reduction of 72.71\% accross different models. An interesting finding is that, for the \texttt{Qwen2.5-14B} and \texttt{Qwen2.5-32B} models, the original SSD offloading system uses the same amount of memory, while the adaptive buffer pool shows slightly higher memory usage for the \texttt{Qwen2.5-32B} model. This occurs because the largest layer, the embedding layer, is identical in both the \texttt{Qwen2.5-14B} and \texttt{Qwen2.5-32B} models, leading to equal memory usage in the original system, which is limited by the largest layer size. However, in the \texttt{Qwen2.5-32B} model, the feedforward-related weights are larger, resulting in slightly increased memory usage compared to the \texttt{Qwen2.5-14B}. This finding highlights that the adaptive buffer pool effectively reveals the actual memory needs of different models.

\paragraph{Fused Overflow Check Mechanism}
\label{sec:evaluation:fused_overflow_check}

To assess the impact of the method described in Section \ref{sec:method:fused_overflow_check}, the analysis focuses on latency as a performance metric and memory efficiency compared to the implementation of the original SSD system. Figure \ref{fig:fused_overflow_check}(a) shows the overflow check latency across various models on Server Configuration 1, which uses an Intel(R) Xeon(R) 6780E CPU. In contrast, Figure \ref{fig:fused_overflow_check}(b) shows the corresponding results for Server Configuration 2, equipped with dual AMD EPYC 7282 CPUs. The latency scales linearly with model size, as the operation examines each gradient for overflow. The fused overflow check mechanism, by removing intermediate tensor creation and multi-step processes, performs a single pass through the values, resulting in an average latency reduction of 97\%. This mechanism runs in every iteration that requires an update, with computational complexity scaling linearly with model size, underscoring the importance of performance optimization. For memory efficiency, Figure \ref{fig:fused_overflow_check_memory_overhead} indicates that MemAscend does not have additional memory overhead compared to ZeRO-Infinity. Unlike the original overflow check, which generates multiple intermediate values and incurs significant memory overhead, the fused mechanism eliminates this overhead by avoiding intermediate computations.

\paragraph{Direct NVMe Engine}

\begin{figure*}[!t]
\centering
\subfloat[Write latency]{\includegraphics[width=1.7in]{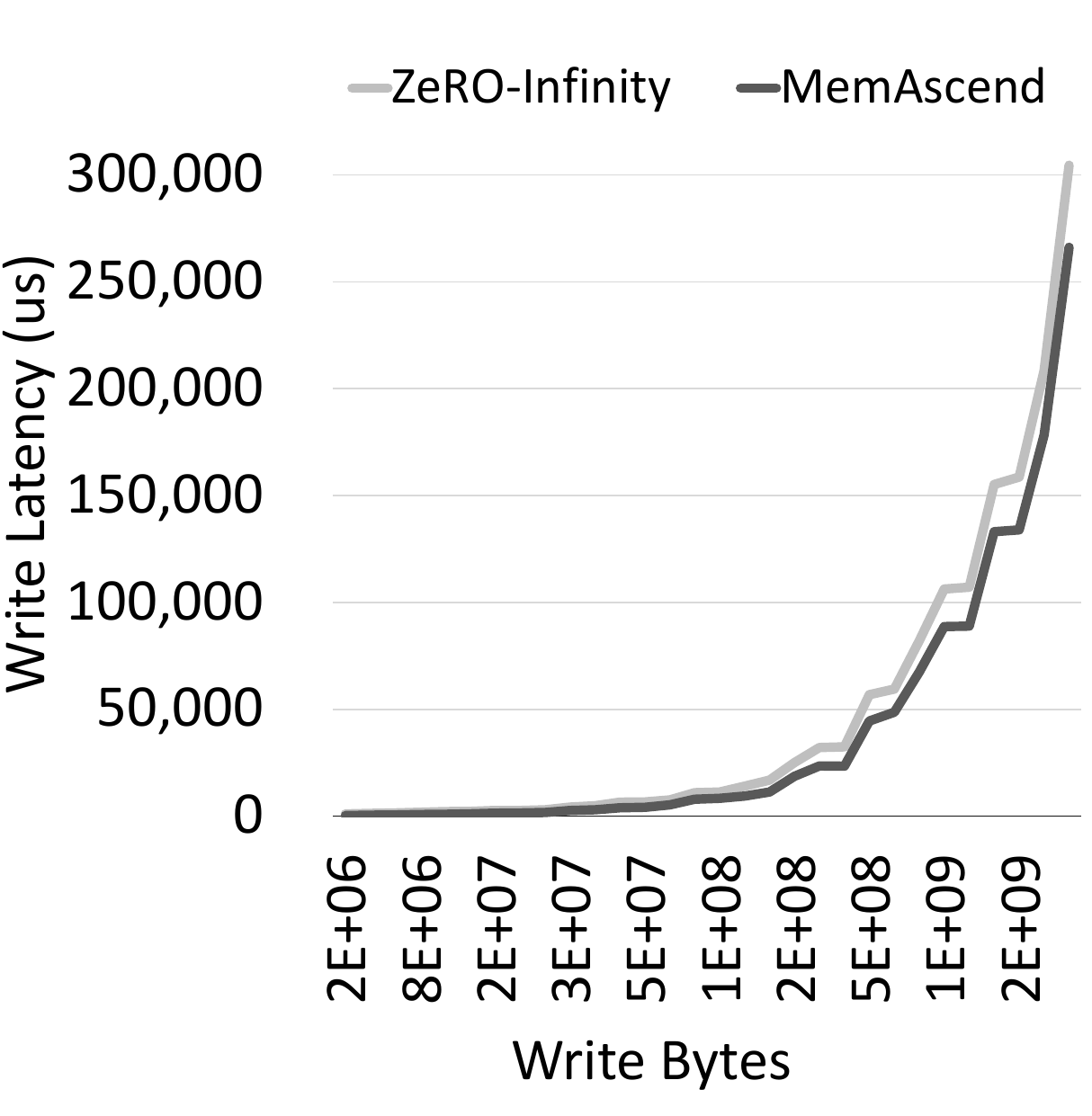}%
\label{fig:write_latency}}
\hfil
\subfloat[Write bandwidth]{\includegraphics[width=1.7in]{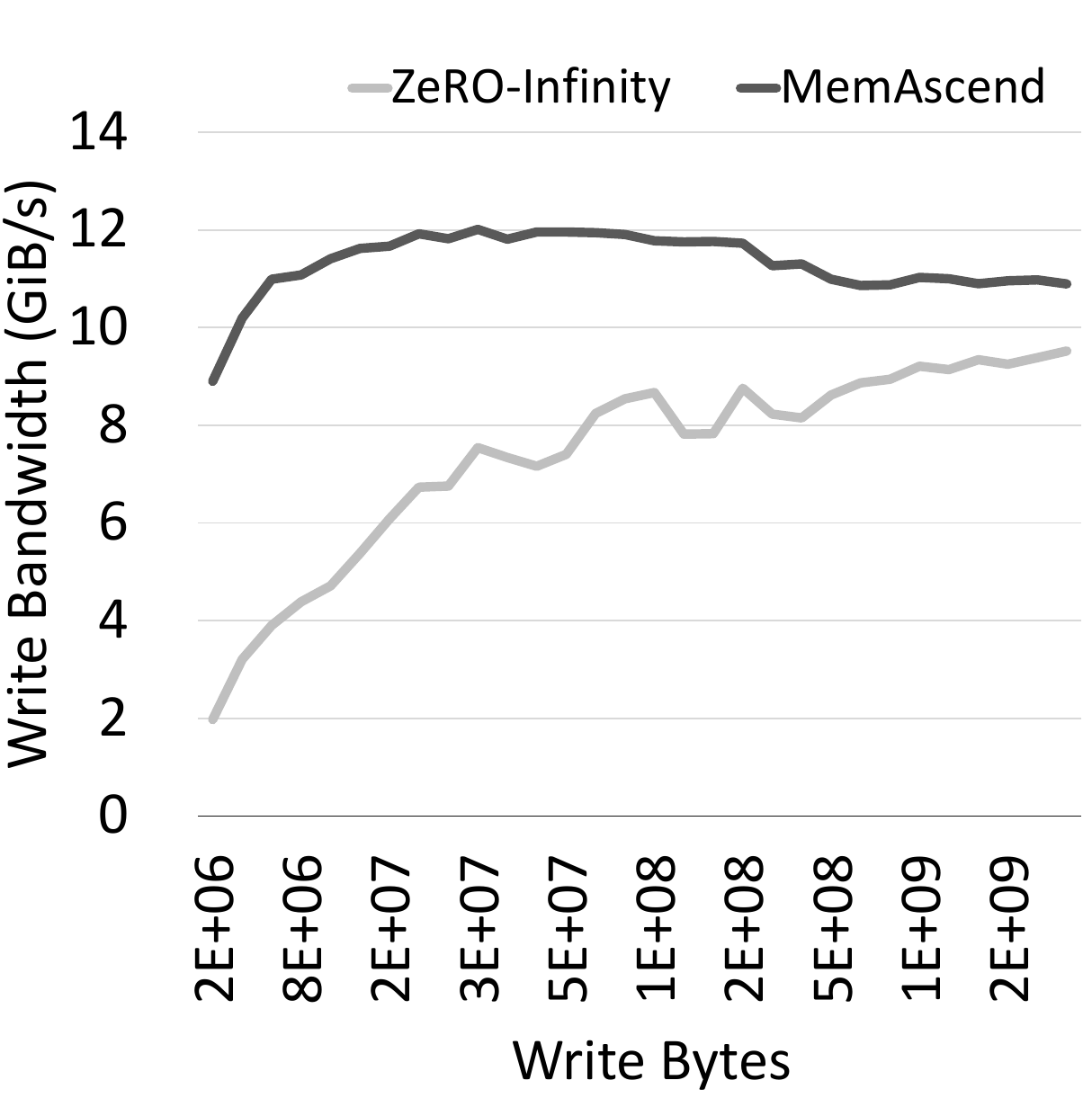}%
\label{fig:write_bandwidth}}
\hfil
\subfloat[Read latency]{\includegraphics[width=1.7in]{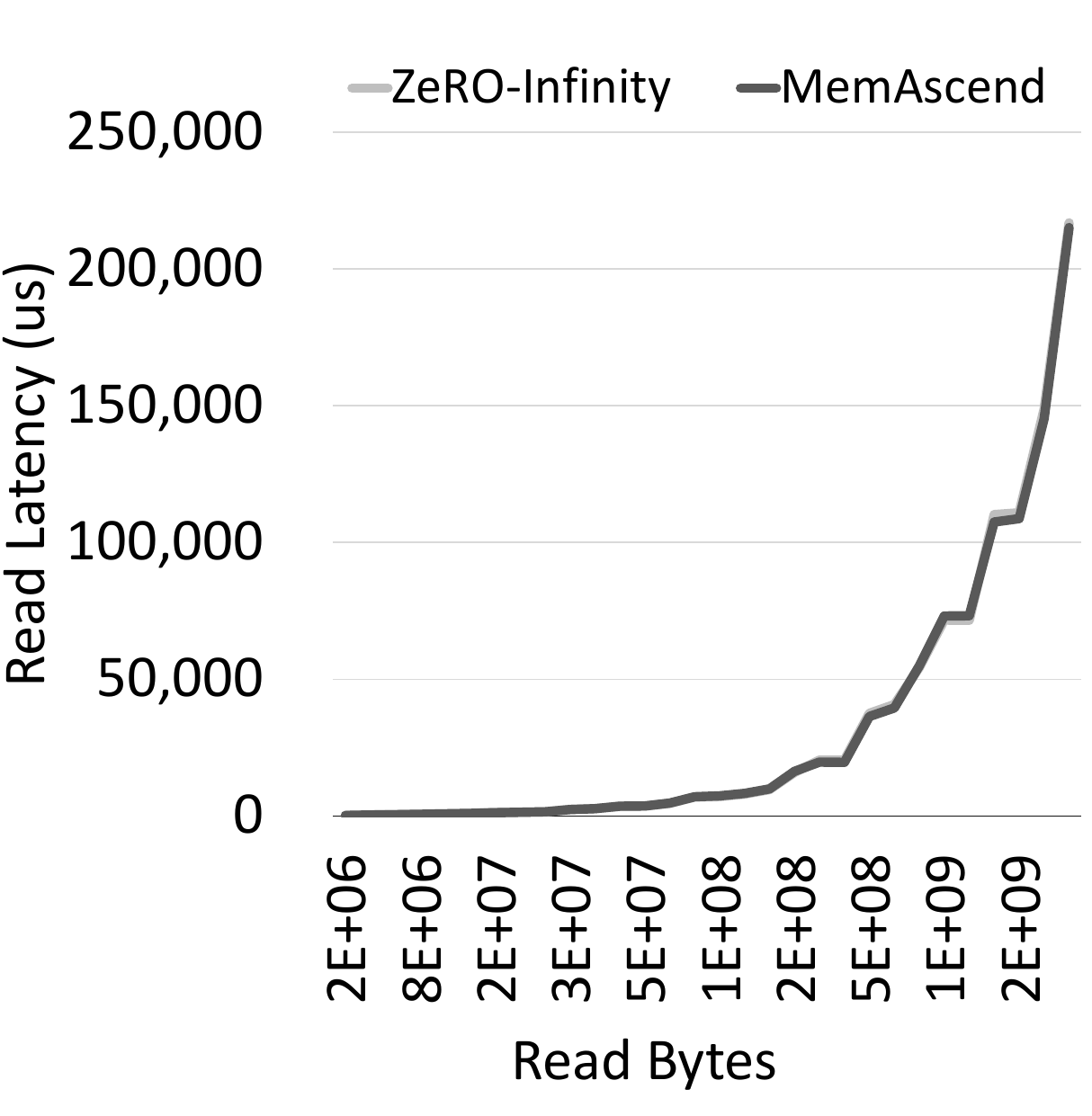}%
\label{fig:read_latency}}
\hfil
\subfloat[Read bandwidth]{\includegraphics[width=1.7in]{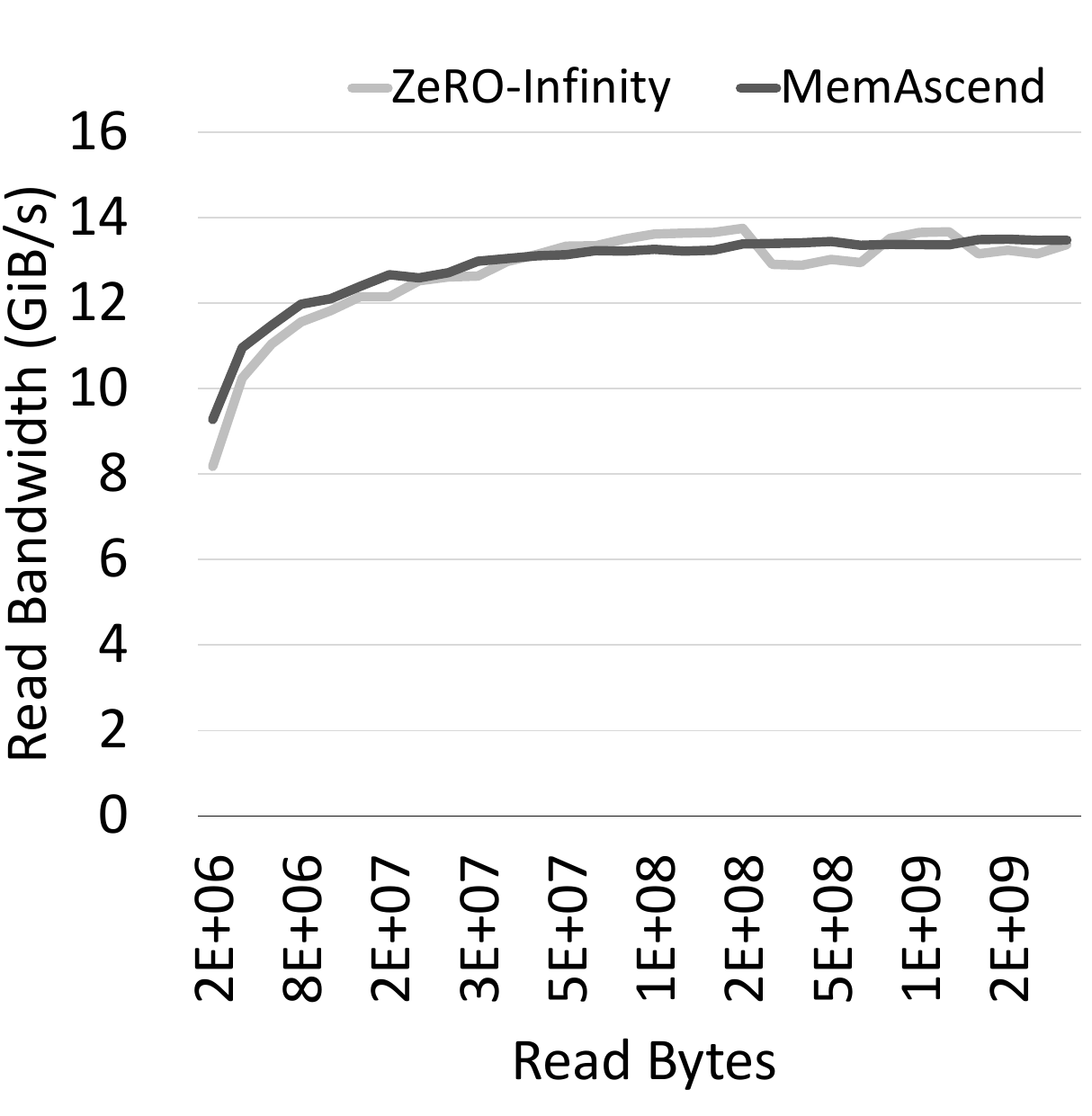}%
\label{fig:read_bandwidth}}
\caption{Comparison of NVMe SSD read/write latency and bandwidth between the ZeRO-Infinity approach and the direct NVMe engine across various tensor sizes.}
\label{fig:direct_nvme_engine}
\vspace{-0.25in}
\end{figure*}

This evaluation examines the performance of the direct NVMe engine, described in Section~\ref{sec:method:direct_nvme_engine}, for SSD I/O in training scenarios with SSD offloading. The engine manages offloading of fp16 and fp32 model weights, gradients, and optimizer states. To determine suitable tensor sizes for benchmarking, model weight distributions were analyzed, excluding tensors that are too small to benefit from offloading. For instance, tensors with fewer than two million elements perform better in CPU memory. Common dense LLM architectures include several layers suitable for SSD offloading, such as feedforward layers, q, k, v, and o projection layers, the embedding layer, and the linear head layer. To ensure a precise comparison with the original SSD offloading technique, tensor size ranges for \texttt{fp16} and \texttt{fp32} data were collected across various models. The analysis also considers multi-GPU setups, where data are split across processes. For example, a 100-million-element tensor divides into 50 million elements per process in a two-GPU configuration. Benchmarks used Configuration 2, as specified in Table~\ref{tab:hardware_specs}. The original SSD offloading system, reliant on filesystem I/O, utilized two SSDs in a RAID0 array formatted with ext4. By contrast, MemAscend employs the eirect NVMe engine, which bypasses software RAID and sends requests directly to SSDs through the NVMe driver.

Figures~\ref{fig:direct_nvme_engine}(a) and~\ref{fig:direct_nvme_engine}(b) demonstrate that the direct NVMe engine significantly lowers write latency and improves write bandwidth across different transfer sizes. For a 2{,}097{,}152-byte tensor, latency decreases from 988.222 microseconds to 219.418 microseconds. For a 3{,}114{,}270{,}720-byte tensor, it decreases from 304{,}604.779 microseconds to 266{,}212.637 microseconds. The bandwidth improvement reaches up to 4.5×, with a 72.04\% average gain. The decreasing trend in MemAscend’s write bandwidth stems from the SSD’s internal DRAM buffers and SLC write cache, which can absorb short bursts. When the transfer size is small and the I/O queue is efficiently utilized, data is served directly from high-speed cache, yielding very high application-observed bandwidth. As the written volume grows, these caches fill, forcing data to be destaged to slower NAND media. The resulting throughput gradually converges toward the sustained rate of the device. In contrast, the filesystem-based baseline performs poorly for small writes due to dominant host-side overheads, then improves as these costs are amortized, leading to the contrasting shapes of the two curves.

\begin{figure}[b]
\centering
\vspace{-0.25in}
\includegraphics[height=1.7in]{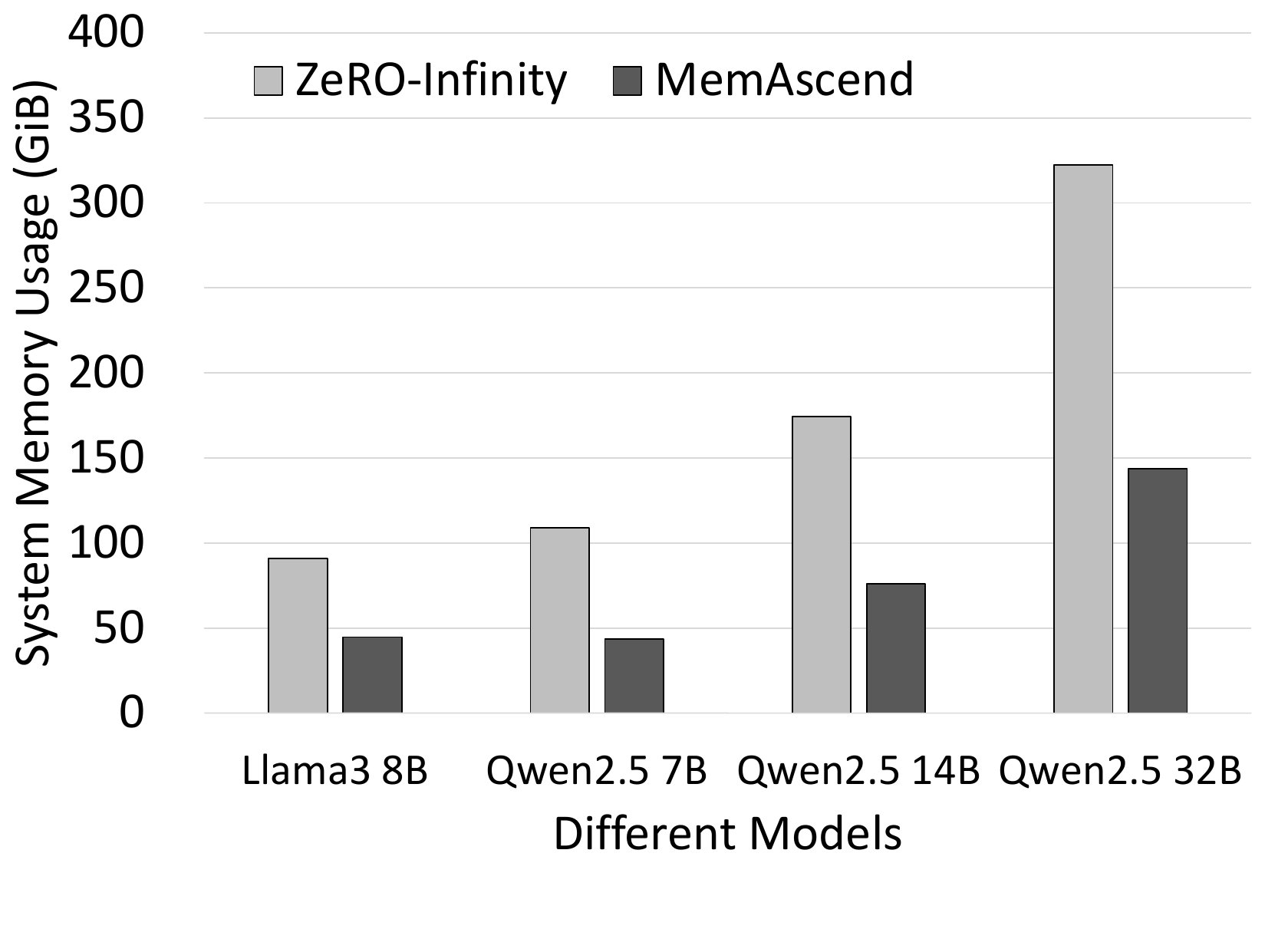}
\vspace{-0.25in}
\caption{Peak system memory usage comparison between ZeRO-Infinity and MemAscend across various models.}
\label{fig:end_to_end_memory}
\end{figure}

The read behavior differs, as shown in Figures~\ref{fig:direct_nvme_engine}(c) and~\ref{fig:direct_nvme_engine}(d). Reads bypass much of the expensive journaling and allocation paths, so both methods achieve similar average read bandwidth. However, MemAscend’s direct LBA I/O avoids pathname resolution, metadata lookups, and RAID-level merges on the critical path, thereby reducing variability. This results in more consistent read performance, even when the mean bandwidth remains comparable to the baseline.

\subsubsection{Assessment of the Integrated Solution}

\paragraph{End-to-End Memory Efficiency}

To evaluate the overall memory efficiency of the proposed MemAscend, a comparison of peak system memory usage against the baseline ZeRO-Infinity system across four model architectures is conducted. This comparison is illustrated in Figure~\ref{fig:end_to_end_memory}, a bar graph designed to highlight the system memory reduction achieved by our approach.
The evaluation results reveal a consistent trend. MemAscend significantly reduces system memory usage across all models tested compared to the baseline. Specifically, for the \texttt{Llama3.1-8B} model, the baseline consumes 91.06 GiB, while our method reduces this to 44.71 GiB, a savings of approximately 50.9\%. For the \texttt{Qwen2.5-7B} model, memory usage drops from 109.06 GiB to 43.67 GiB, leading to a reduction of 60.0\%. In the \texttt{Qwen2.5-14B} case, the baseline requires 174.5 GiB, whereas MemAscend uses only 76.1 GiB, achieving a 56.4\% decrease. Finally, for the \texttt{Qwen2.5-32B} model, the baseline demands 322.3 GiB, while our method lowers this to 143.6 GiB, resulting in a reduction of 55.4\%. On average, MemAscend achieves a memory reduction of approximately 55.7\% across these models. This substantial memory savings underscores the effectiveness of our approach in optimizing resource utilization, with free memory allowing for further enhancements, such as supporting a longer context by utilizing the released space for activation checkpoint values, as discussed in the following sections. The effectiveness of MemAscend is the result of a combination of optimizations detailed in Section~\ref{sec:analysis:memory_efficiency}. 

\paragraph{Enabling Longer Context}

\begin{figure*}[!t]
\centering
\vspace{-0.2in}
\subfloat[Llama3.1-8B]{\includegraphics[width=1.7in]{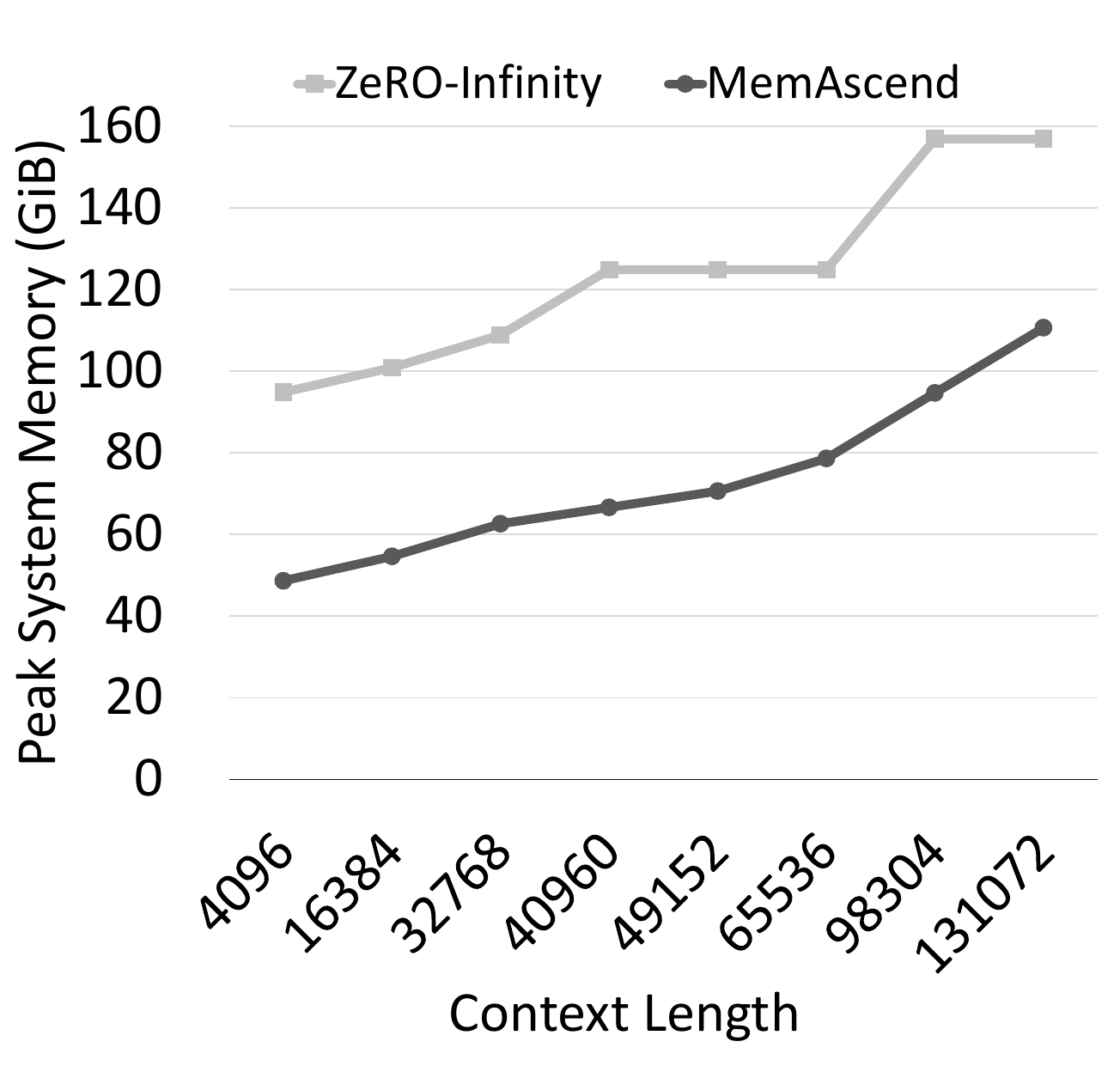}%
\label{fig:context_length_8B}}
\hfil
\subfloat[Qwen2.5-7B]{\includegraphics[width=1.7in]{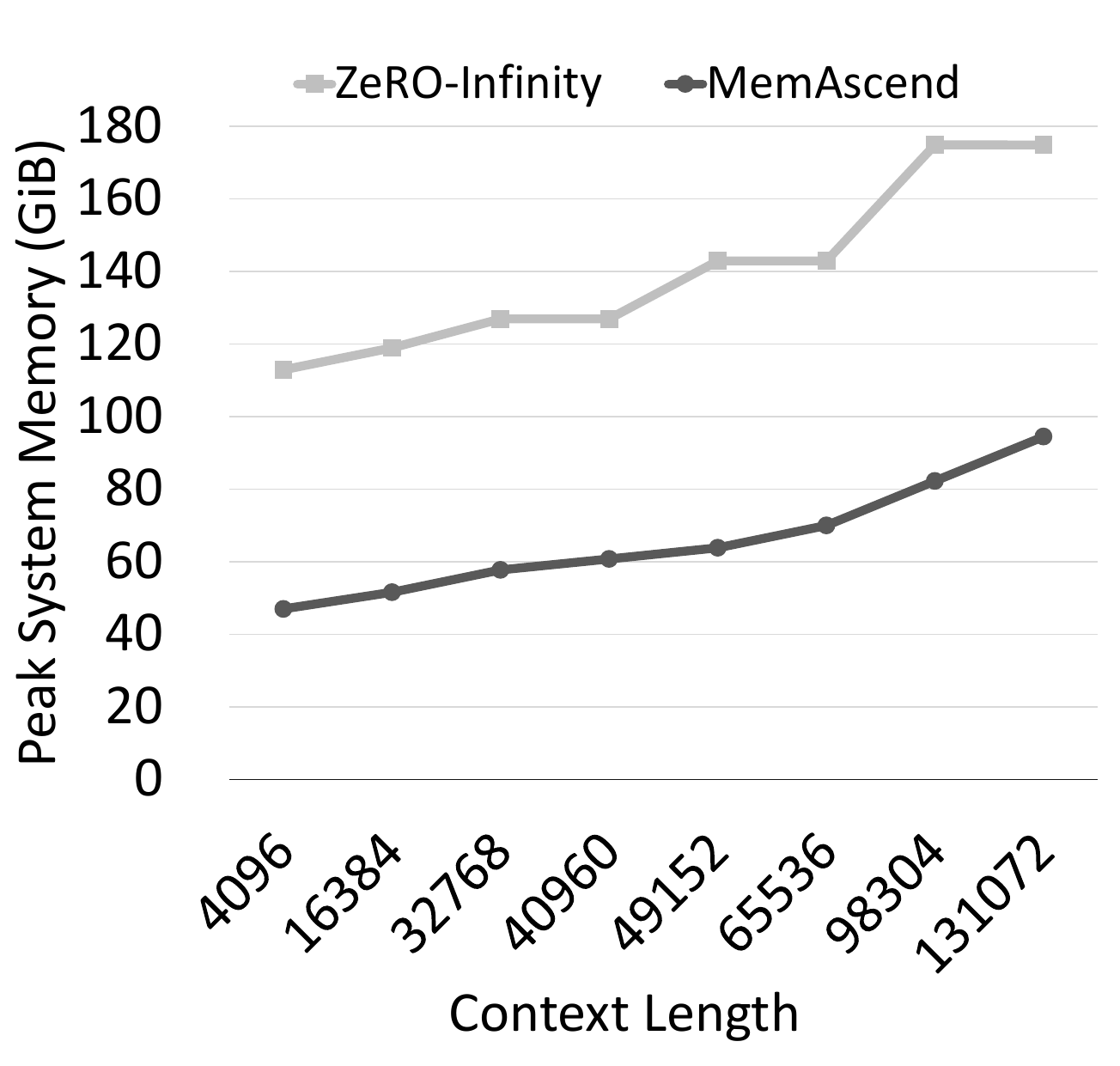}%
\label{fig:context_length_7B}}
\hfil
\subfloat[Qwen2.5-14B]{\includegraphics[width=1.7in]{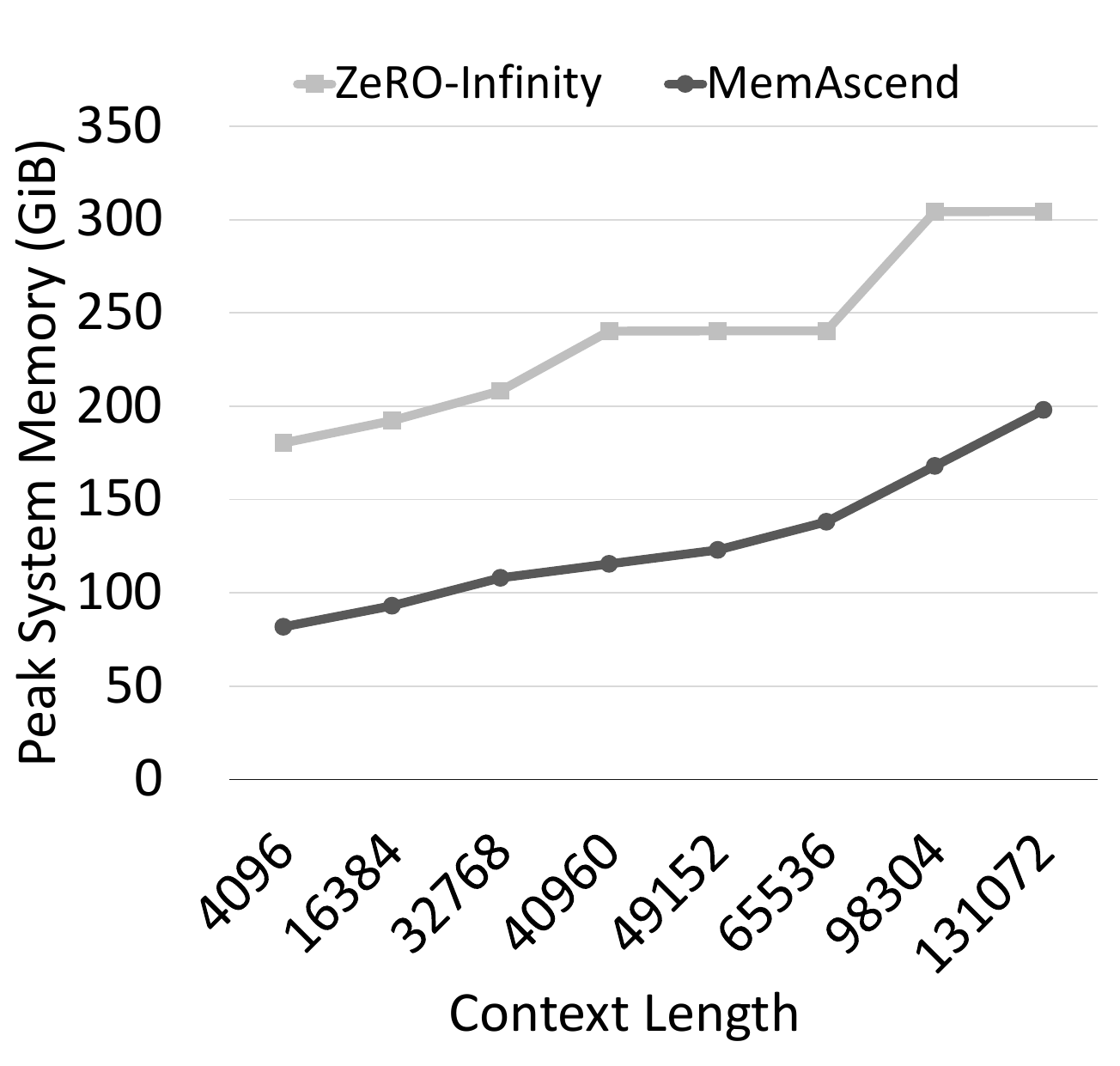}%
\label{fig:context_length_14B}}
\hfil
\subfloat[Qwen2.5-32B]{\includegraphics[width=1.7in]{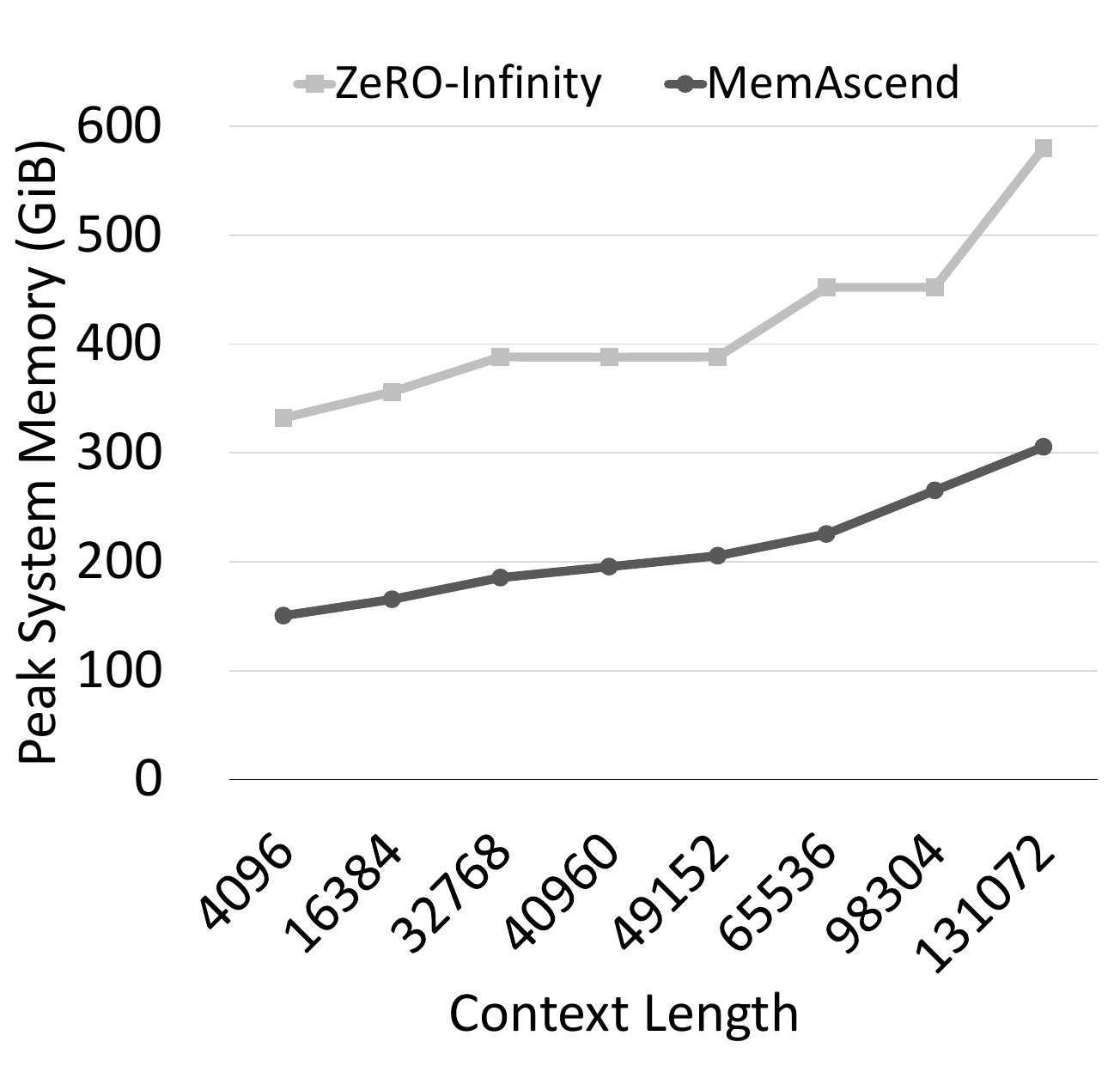}%
\label{fig:context_length_32B}}
\caption{Comparison of peak system memory usage between ZeRO-Infinity and MemAscend in a 2-GPU setup, with context lengths ranging from 4,096 to 131,072 tokens. The results demonstrate MemAscend ability to lower memory consumption significantly, allowing training with longer context lengths that ZeRO-Infinity cannot support within the same hardware constraints.}
\vspace{-0.2in}
\label{fig:longer_context}
\end{figure*}

\begin{figure*}[!t]
\centering
\subfloat[Llama3.1-8B]{\includegraphics[width=1.7in]{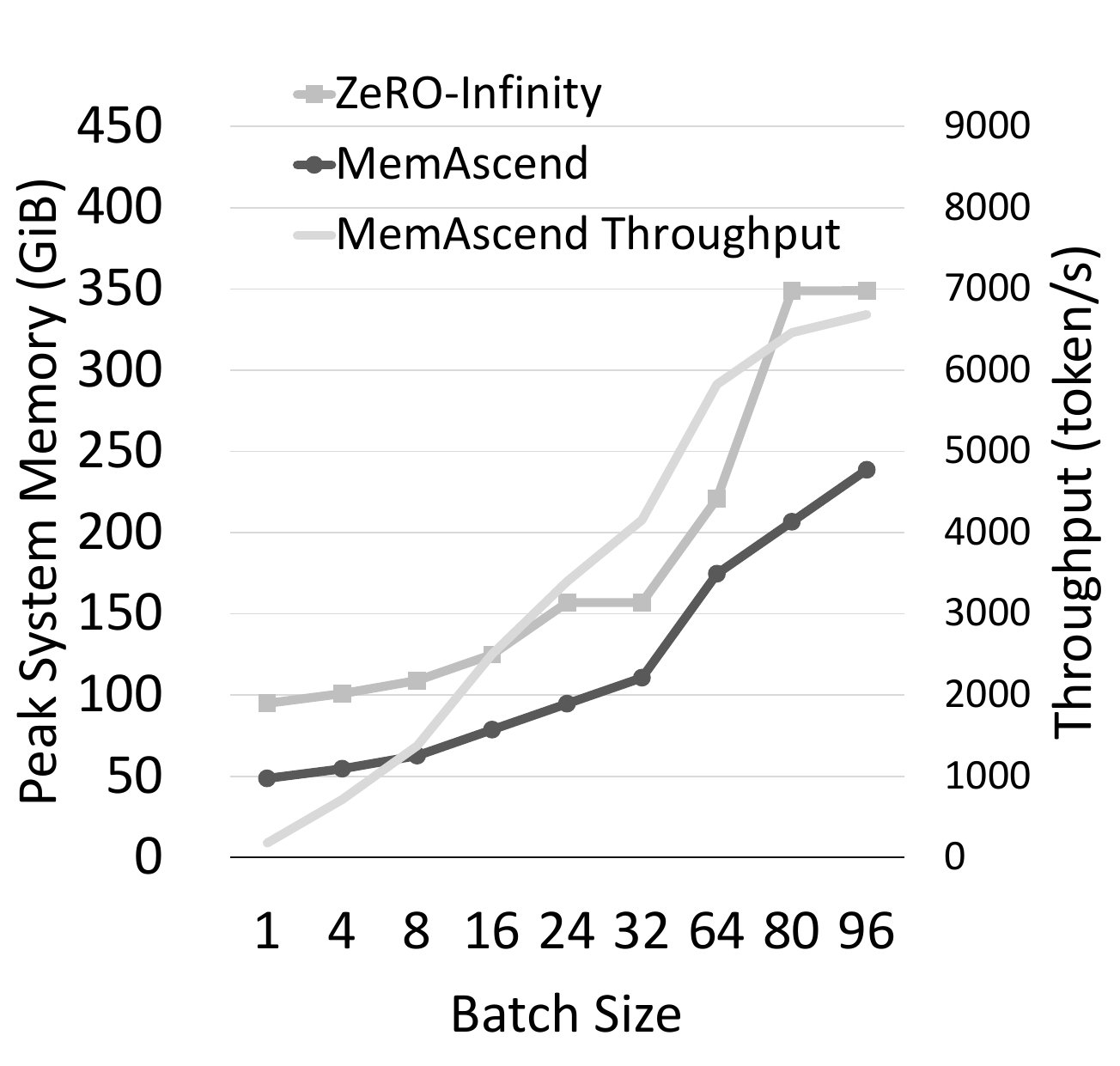}%
}
\hfil
\subfloat[Qwen2.5-7B]{\includegraphics[width=1.7in]{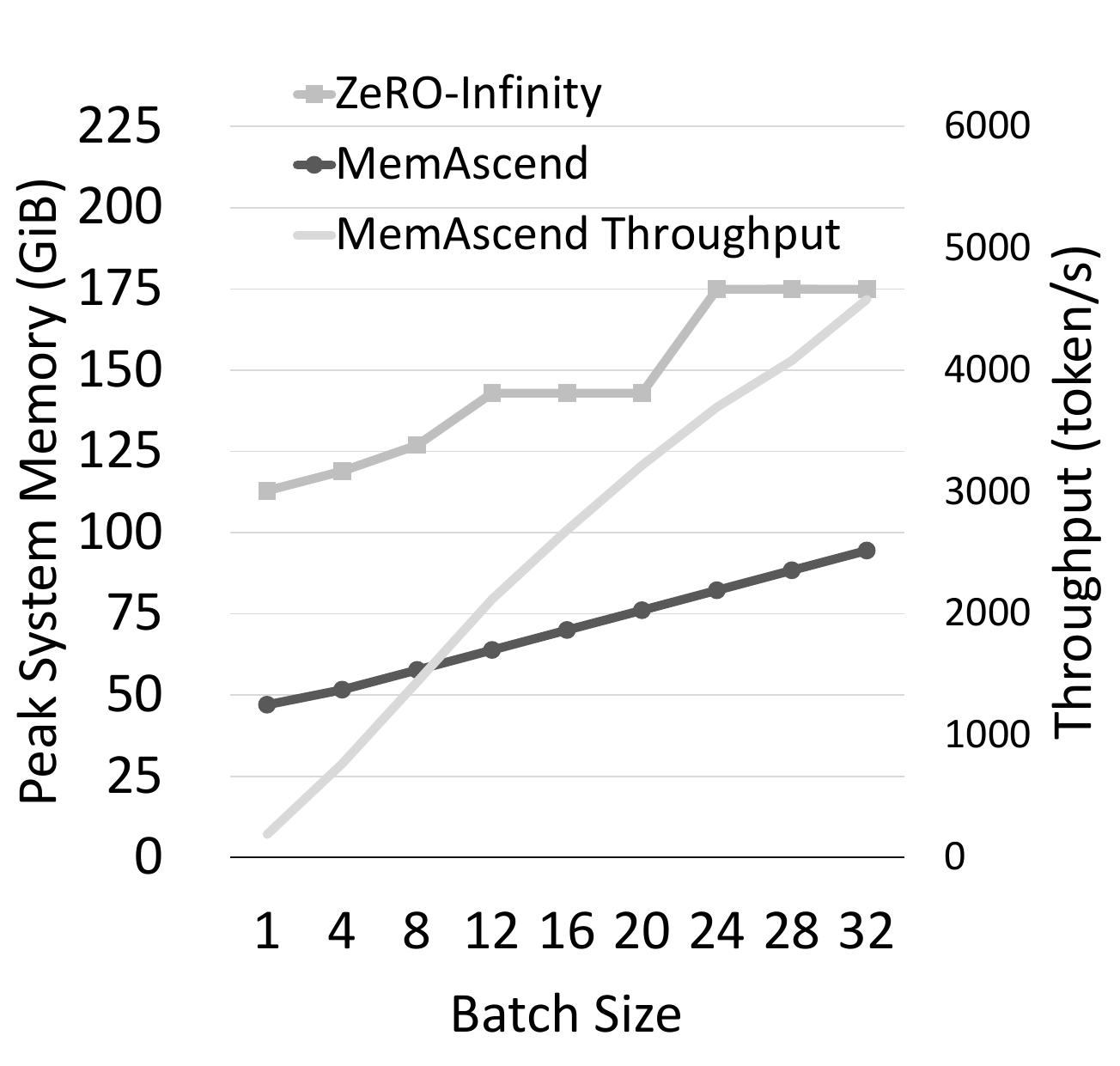}%
}
\hfil
\subfloat[Qwen2.5-14B]{\includegraphics[width=1.7in]{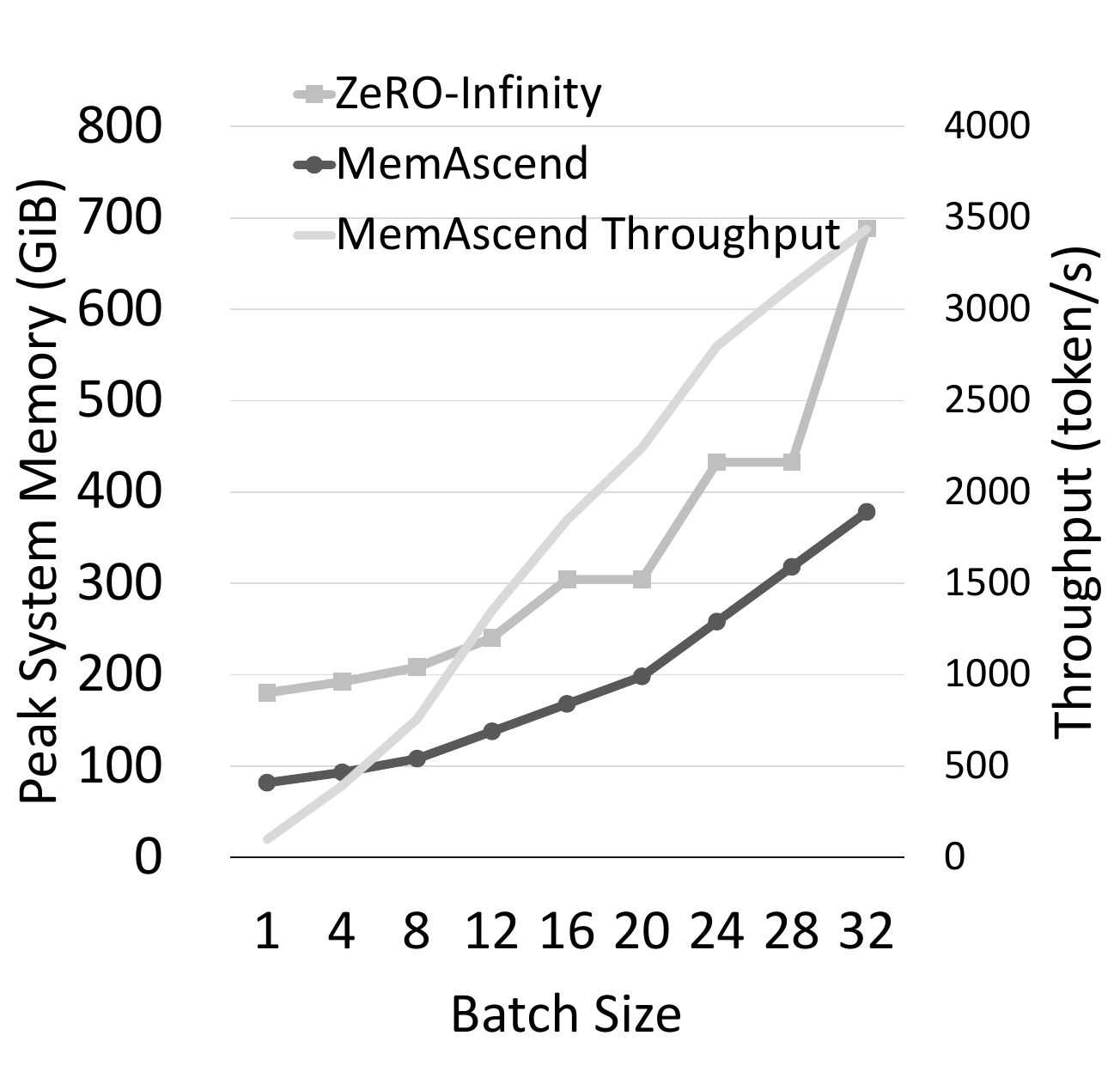}%
}
\hfil
\subfloat[Qwen2.5-32B]{\includegraphics[width=1.7in]{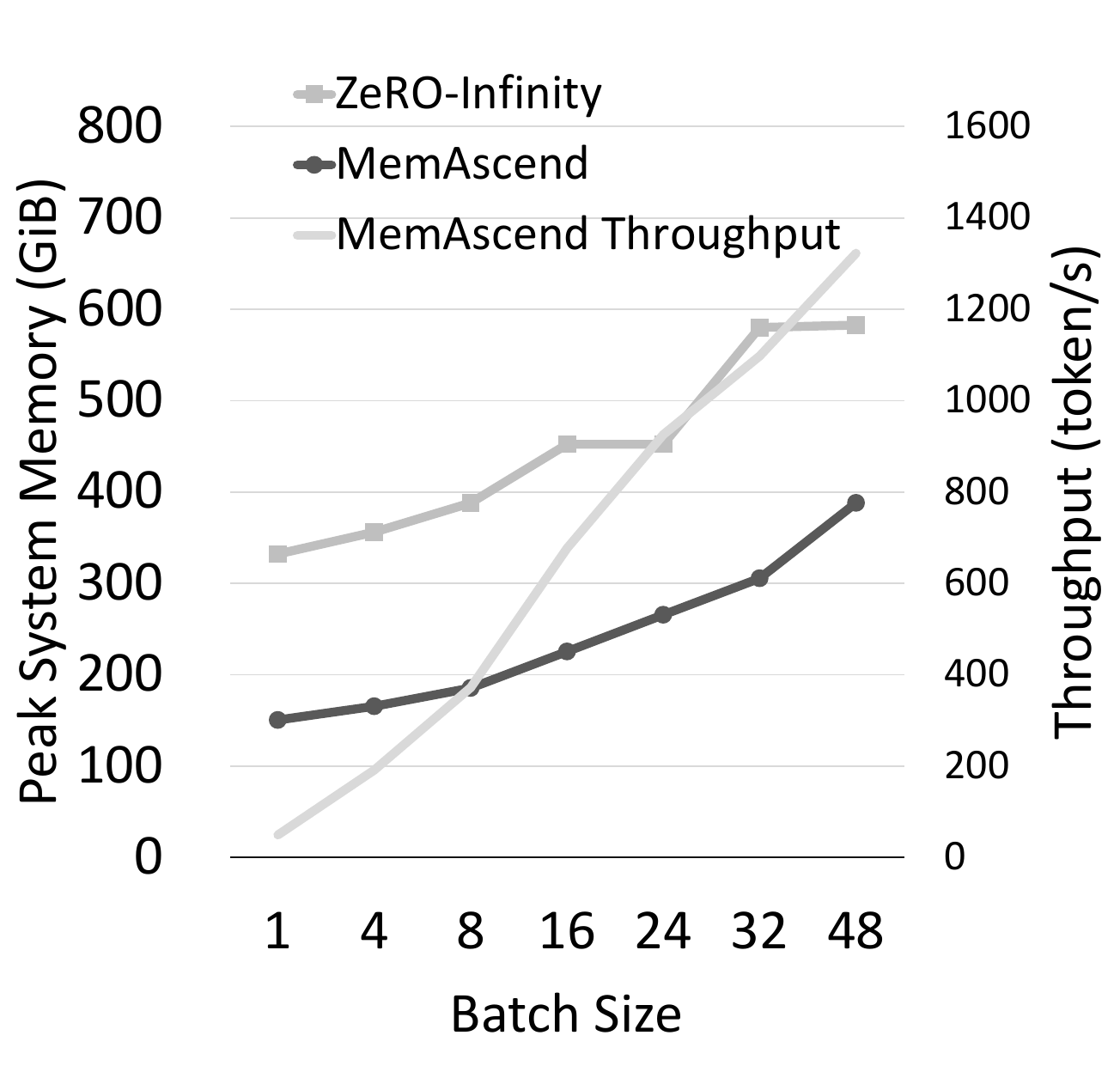}%
}
\caption{System memory usage and throughput are compared between ZeRO-Infinity and MemAscend across varying batch sizes, using a fixed context length of 4,096 tokens on a 2-GPU setup with NVIDIA H100s. MemAscend maintained lower system memory usage across all batch sizes, enabling efficient scaling to larger batches. This allowed the system to reach saturated throughput without breaching memory constraints, while ZeRO-Infinity exhibited earlier memory bottlenecks that limited batch size and overall performance.}
\vspace{-0.2in}
\label{fig:evaluations_higher_performance}
\end{figure*}

Figure~\ref{fig:longer_context} illustrates the peak system memory usage in various context lengths for four compared LLMs under a 2-GPU setup
. In each figure, the x-axis represents the context length, ranging from 4,096 to 131,072 tokens, while the y-axis denotes system memory usage in GiB. These figures aim to demonstrate how our method reduces memory consumption, enabling support for longer context lengths under the same hardware configuration. The general trend across all figures reveals that system memory usage increases with context length for both the baseline and our method. However, MemAscend consistently exhibits lower memory usage and a slower scaling rate compared to ZeRO-Infinity. For example, in Figure~\ref{fig:longer_context}(a) for \texttt{Llama3.1-8B}, the baseline memory usage increases from 94.88 GiB at 4,096 tokens to 156.88 GiB at 131,072 tokens, while our method increases from 48.67 GiB to 110.67 GiB, achieving an average reduction of 41.65\%. Similarly, in Figure~\ref{fig:longer_context}(d) for \texttt{Qwen2.5-32B}, the baseline jumps from 332.12 GiB to 580.12 GiB, while MemAscend scales from 150.51 GiB to 305.5 GiB, achieving an average reduction of 49.48\%. The effectiveness of MemAscend is the result of a combination of optimizations detailed in Section~\ref{sec:analysis:enable_longer_context}. 

\paragraph{Enabling Higher Throughput with Reclaimed System Memory}

To evaluate how MemAscend enhances performance by leveraging memory savings, Figure~\ref{fig:evaluations_higher_performance} is presented and is composed of four subfigures to illustrate memory usage and throughput across different batch sizes for the four compared LLMs. Each subfigure corresponds to one model, with the x-axis representing batch size (ranging from 1 to 96, depending on the model), the left y-axis indicating system memory usage in GiB, and the right y-axis showing throughput in tokens per second (tokens/s). In each subfigure, three lines are plotted: one for throughput trends, one for baseline system memory usage, and one for the proposed MemAscend system memory usage. These results were obtained using Configuration 1 (2 NVIDIA H100 GPUs) with a fixed context length of 4,096 tokens. The general trend observed in all the subfigures is two-fold. First, system memory usage increases with batch size for both the baseline and MemAscend, reflecting the growing demand for activation checkpointed values as batch size scales (see Equation~\ref{eq:activation_checkpoint}). However, MemAscend consistently consumes significantly less memory than the baseline across all models and batch sizes. 

For example, in the \texttt{Llama3.1-8B} subfigure, the baseline memory usage rises from 94.89 GiB at batch size 1 to 348.91 GiB at batch size 96, while MemAscend increases from 48.67 GiB to 238.69 GiB. Similar patterns are evident in the \texttt{Qwen2.5} models, with memory savings becoming more pronounced at larger batch sizes. Across all models, MemAscend achieves an average memory reduction of approximately 42.8\%, enabling support for larger batch sizes under the same system memory constraints. Second, throughput increases near-linearly with batch size for all models, demonstrating that larger batch sizes enhance GPU utilization. For example, in the \texttt{Qwen2.5-32B} subfigure, the throughput grows from 49.32 tokens/s at batch size 1 to 1,322.55 tokens/s at batch size 48, representing a 26.82× improvement, which highlights the performance potential unlocked by larger batch sizes. These results show that the proposed MemAscend can achieve higher throughput under identical configurations by better utilizing GPU resources through larger batch sizes. The effectiveness of MemAscend is due to its ability to optimize system memory usage, as detailed in Section~\ref{sec:analysis:enable_large_batch_size}. 

\paragraph{End-to-End Performance Improvement}

\begin{table}[!t]
  \centering
  \caption{End-to-End performance improvement from ZeRO-Infinity to MemAscend for Configurations 1 (C1) and 2 (C2)}
  \vspace{-0.1in}
  \label{tab:end_to_end_performance}
  \resizebox{\columnwidth}{!}{%
    \begin{tabular}{|c|c|c|c|}
      \hline
      \textbf{Model Name} & \textbf{Batch (C1 / C2)} & \textbf{C1 Improvement (\%)} & \textbf{C2 Improvement (\%)} \\ \hline
      Llama3.1-8B & 8 / 8   & 6.97 & 12.91 \\ \hline
      Llama3.1-8B & 80 / 20 & 2.72 & 7.52 \\ \hline
      Qwen2.5-7B  & 8 / 8   & 5.53 & 14.02 \\ \hline
      Qwen2.5-7B  & 64 / 20 & 3.73 & 8.36  \\ \hline
      Qwen2.5-14B & 8 / 4   & 6.45 & 18.86 \\ \hline
      Qwen2.5-14B & 64 / 16 & 3.28 & 6.77 \\ \hline
      Qwen2.5-32B & 8 / 4   & 5.64 & 18.43 \\ \hline
      Qwen2.5-32B & 48 / 8  & 2.89 & 16.42 \\ \hline
    \end{tabular}%
  }
  \vspace{-0.3in}
\end{table}

The end-to-end performance of MemAscend is evaluated by comparing its training throughput against the ZeRO-Infinity baseline under two hardware configurations described in Table~\ref{tab:hardware_specs}. Both systems were tested with the direct NVMe engine enabled. ZeRO-Infinity without this engine, which relies on a traditional filesystem-based approach, is unstable and prone to hanging, so it was excluded from this comparison. Table~\ref{tab:end_to_end_performance} reports the percentages of improvement in performance in a range of models, batch sizes, and hardware configurations, along with the gains achieved by MemAscend over ZeRO-Infinity. All tests used a fixed context length of 4096. For each configuration, we include several batch sizes to capture workload variation. The table lists the model, the batch sizes used for Configurations 1 and 2, and the corresponding percentage improvements.

In both configurations, MemAscend consistently outperformed ZeRO-Infinity, with improvements ranging from 2.72\% to 6.97\% in Configuration 1. In Configuration 2, the gains were more substantial, ranging from 6.77\% to 18.86\%. This difference underscores a key observation: MemAscend advantages are more pronounced when the baseline suffers from higher latency in overflow checking, as in Configuration 2, where slower CPU performance increases the cost of overflow checking. Notably, performance gains tend to be larger at smaller batch sizes. For example, \texttt{Qwen2.5-14B} shows an improvement of 18. 86\% at batch size 4 in Configuration 2 and 6.77\% at batch size 16. This trend suggests that when batch sizes are smaller and forward and backward pass times are reduced, the relative impact of our optimizations becomes more significant. In summary, the proposed MemAscend not only reduces system memory usage but also achieves higher throughput and lower latency than the original SSD offloading technique.

\paragraph{Performance Across Diverse Model Architectures}

\begin{table}[h]
\vspace{-0.2in}
\caption{Hardware Specifications for Sparse Model (MoE)}
\vspace{-0.1in}
\label{tab:hardware_specs_moe}
\footnotesize
\centering
\resizebox{2.8in}{!}{
\begin{tabular}{|c||c|}
    \hline
    \textbf{Component} & \textbf{Configuration 3} \\
    \hline
    CPU & Intel Xeon 8480+ \\
    \hline
    CPU Memory & 1 TB 4800 MT/s DDR5 \\
    \hline
    PCIe & Gen5 \\
    \hline
    GPU & 2 $\times$ NVIDIA H100 SXM5 (with NVL) \\
    \hline
    SSD & 2 $\times$ Samsung 980 Pro (1 TB each) \\
    \hline
\end{tabular}
}
\vspace{-0.1in}
\end{table}

\begin{figure}[h]
\centering
\vspace{-0.25in}
\subfloat[Different Context Length]{\includegraphics[height=1.4in]{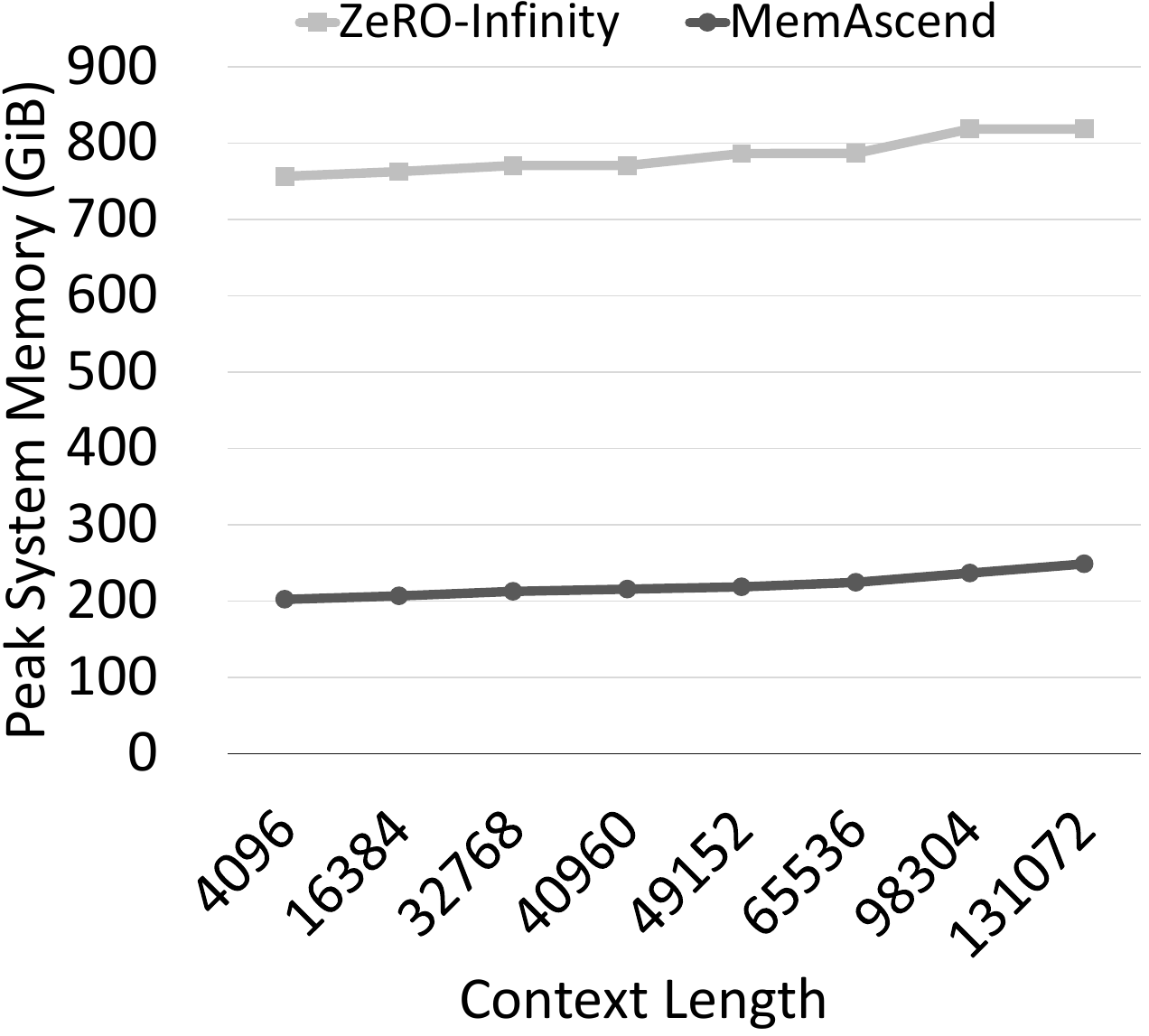}%
}
\hfil
\subfloat[Different Batch Size]{\includegraphics[height=1.4in]{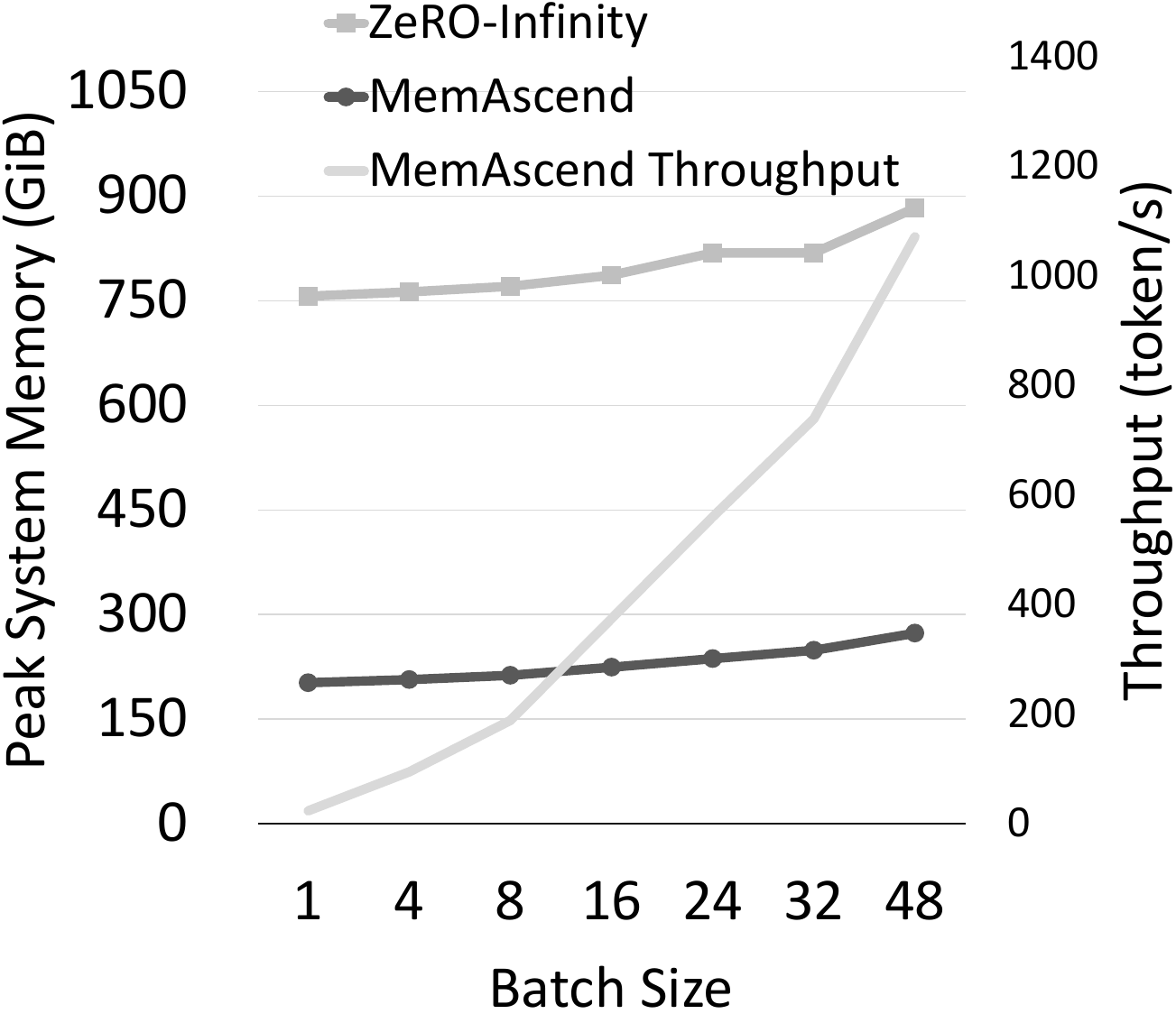}%
}
\caption{Comparison of system memory usage and throughput between ZeRO-Infinity and MemAscend across different context lengths (batch size = 1) and batch sizes (context length = 4096).}
\vspace{-0.15in}
\label{fig:evaluations_qwenmoe}
\end{figure}

To demonstrate the memory efficiency of MemAscend on diverse model architectures, this section evaluates its performance on sparse models, also known as the Mixture-of-Experts (MoE) model (Configuration 3 in Table~\ref{tab:hardware_specs_moe}). Figures~\ref{fig:evaluations_qwenmoe}(a) and \ref{fig:evaluations_qwenmoe}(b) show that baseline memory usage grows from 756.73~GiB at 4{,}096 tokens to 818.74~GiB at 131{,}072 tokens, whereas MemAscend increases only from 202.24~GiB to 248.75~GiB, achieving average reductions of 71.87\% and 71.40\%, respectively. MoE architectures introduce many small expert feed-forward modules (up to 128 for \texttt{Qwen3-30B-A3B}), increasing the number of buffers needed for SSD offloading. MemAscend’s adaptive buffer pool mitigates this pressure by allocating only the exact buffer size required for each expert. In contrast, the baseline allocator must size all buffers to match the largest layer (e.g., embeddings), resulting in significant unnecessary overhead. These results confirm that MemAscend maintains strong memory efficiency and robustness across both dense and sparse model families.


\paragraph{Convergence Behavior}

\begin{figure}[!t]
\centering
\subfloat[Training]{\includegraphics[height=1.35in]{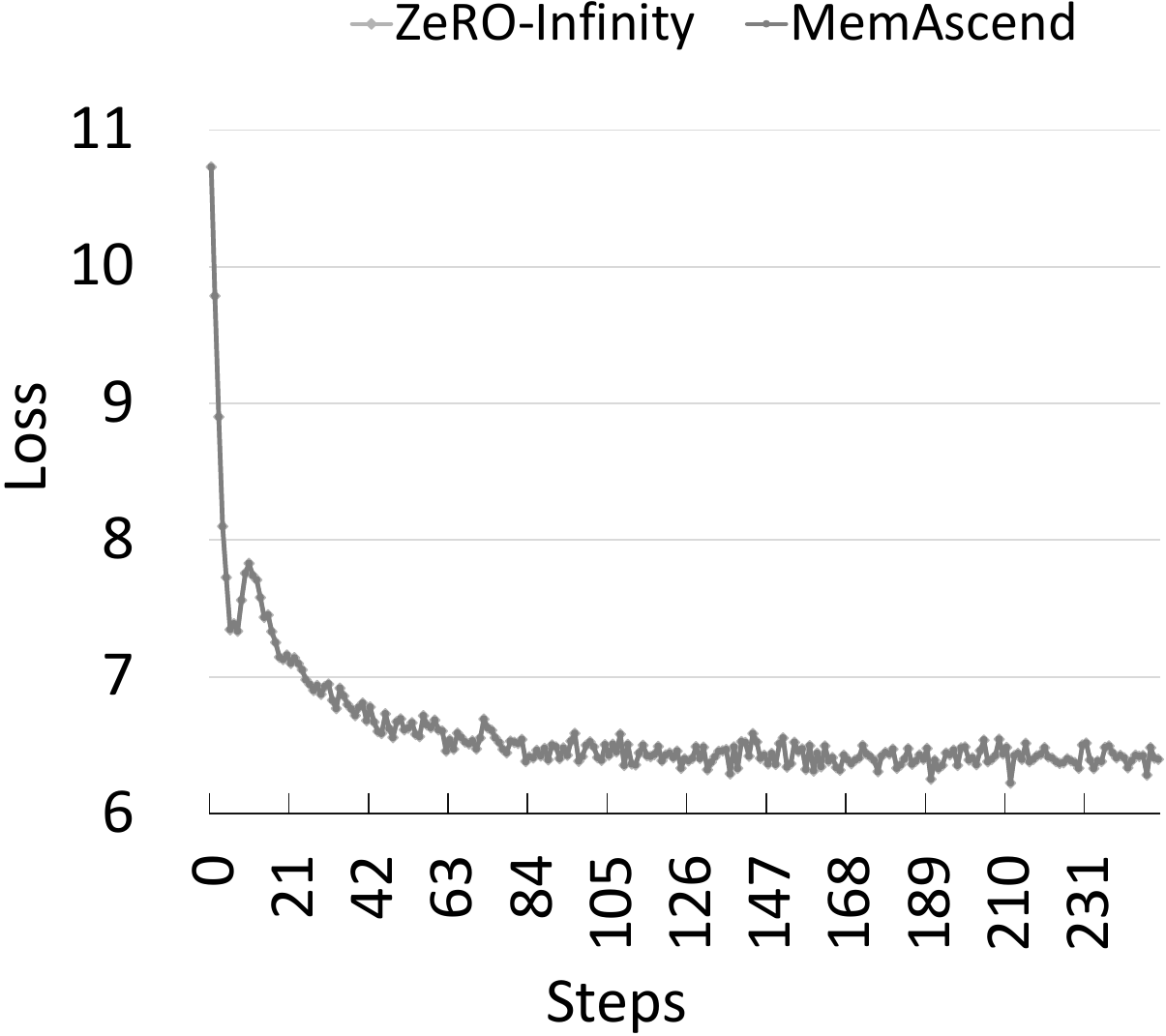}%
}
\hfil
\subfloat[Validation]{\includegraphics[height=1.35in]{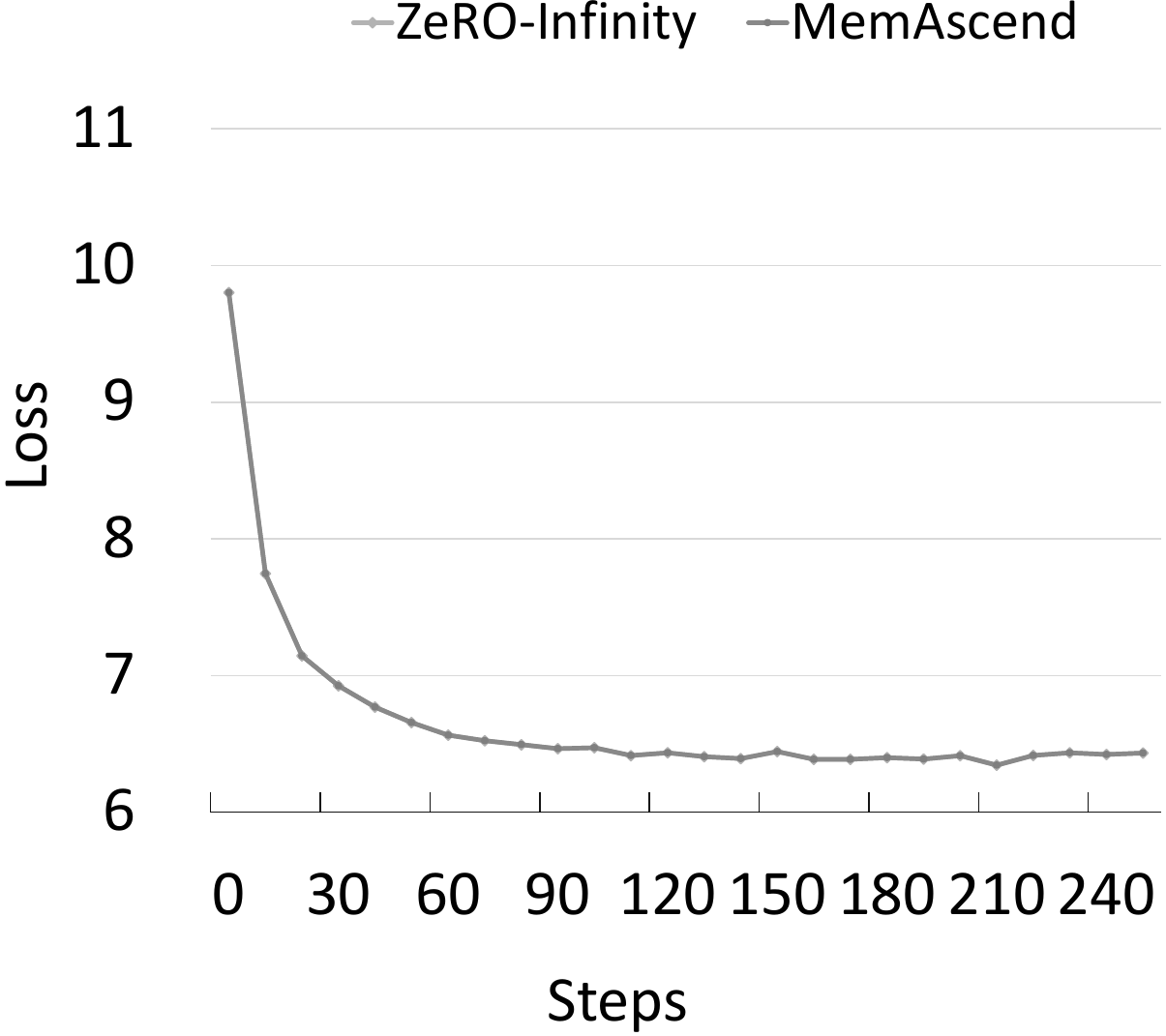}%
}
\caption{Comparison of loss convergence behavior between ZeRO-Infinity and MemAscend.}
\vspace{-0.25in}
\label{fig:evaluations_convergence}
\end{figure}

To understand whether the proposed MemAscend changes convergence behavior, we compare its loss trajectory against ZeRO-Infinity using a real training workload. MemAscend introduces only system-level optimizations—buffer management, pinned-memory allocation, fused overflow checking, and direct NVMe access—without modifying numerical kernels. Therefore, no accuracy deviation is expected. This study evaluate \texttt{Qwen2.5-0.5B} on the \texttt{OpenWebText} dataset~\cite{Gokaslan2019OpenWeb} with batch size 128 and sequence length 1024 for 250 training steps. Figures~\ref{fig:evaluations_convergence}(a) and~\ref{fig:evaluations_convergence}(b) plot the training and evaluation loss curves, respectively. As shown, MemAscend exhibits identical convergence behavior to ZeRO-Infinity throughout the entire training trajectory, confirming that our framework fully preserves model accuracy.

\subsubsection{Extended Experiments}
This section presents additional experiments that fall outside the main scope of the paper. The first explores the use of \texttt{bf16} precision for optimizer-related values. This setup differs from the baseline, which relies on \texttt{fp32} for optimizer-related values as part of \texttt{fp16} mixed-precision training with SSD offloading. Because of the difference in numerical precision, the loss convergence behavior will be different compared to ZeRO-Infinity. Nonetheless, enabling \texttt{bf16} for optimizer states shows promising improvements, especially in reducing I/O load in SSD offloading systems. The second subsection examines mixed-precision training with \texttt{bf16} instead of the \texttt{fp16}-based setup used in the main experiments. This test demonstrates the method's adaptability and effectiveness in a \texttt{bf16} training environment. Together, these experiments highlight MemAscend's flexibility and stability across different precisions.


\paragraph{bf16 Half-Precision Optimizer}

\begin{figure}[!t]
    \centering
    \begin{minipage}{0.24\textwidth}
        \centering
        \includegraphics[height=1.5in]{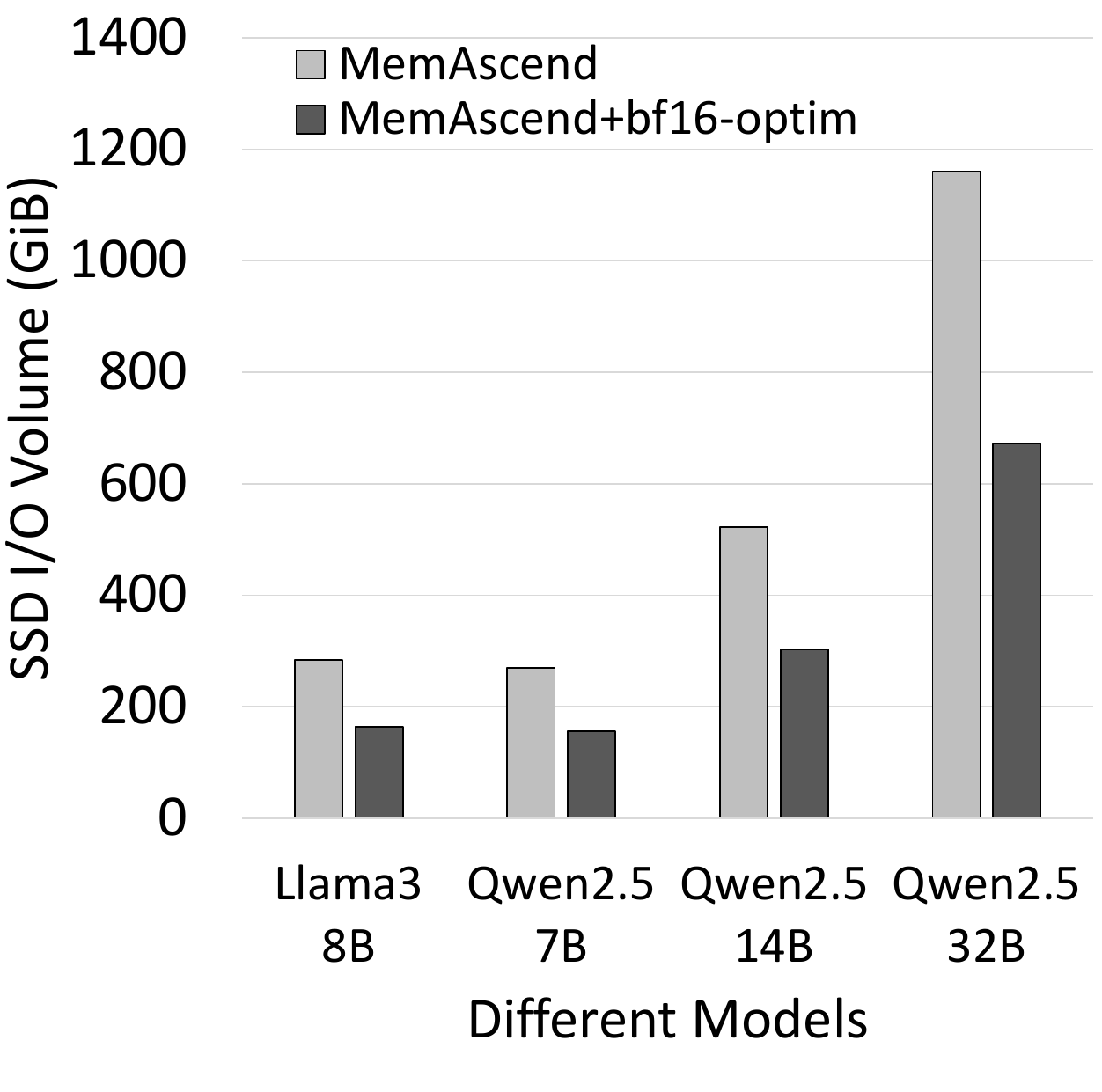}
        \vspace{-0.15in}
        \caption{Comparison of total I/O volume per iteration for MemAscend and MemAscend with bf16 optimizer across various models.}
        \label{fig:bf16_io_footprint}
        \vspace{-0.15in}
    \end{minipage}%
    \hfill
    \begin{minipage}{0.24\textwidth}
        \centering
        \includegraphics[height=1.5in]{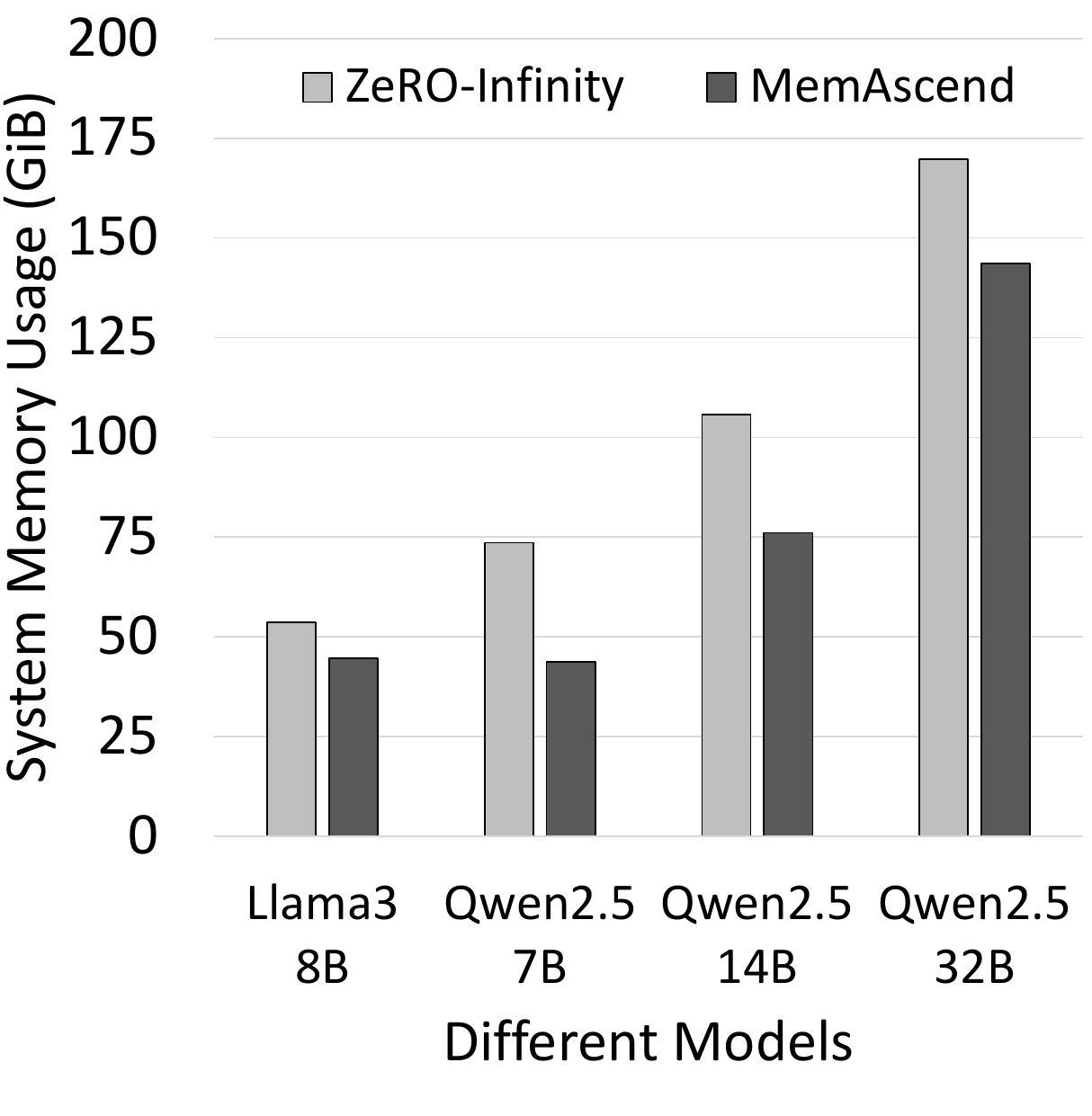}
        \vspace{-0.15in}
        \caption{Comparison of peak system memory usage for ZeRO-Infinity and MemAscend in bf16 mixed-precision training across models.}
        \label{fig:bf16_mixed_precision_mem}
        \vspace{-0.15in}
    \end{minipage}%
\end{figure}

The original baseline supported only \texttt{fp32} optimizer states. Recent work has investigated lower-precision alternatives, including 8-bit \cite{dettmers20228bitoptimizersblockwisequantization} and 4-bit \cite{li2023memoryefficientoptimizers4bit} formats. These reduced-precision representations are especially useful in SSD offloading systems, where lowering the I/O demand is critical. To explore this benefit, \texttt{bf16} optimizer support is integrated into MemAscend. Notably, \texttt{bf16} is selected for optimizer state compression because it incurs minimal numerical-precision overhead and requires only direct truncation from \texttt{fp32}. More aggressive formats (e.g., \texttt{fp8}, 8-bit or 4-bit quantization) require additional scaling, dynamic-range control, and overflow handling, which would introduce added computational complexity not central to the focus of this study. Then, the I/O volume per iteration (Figure~\ref{fig:bf16_io_footprint}) and end-to-end throughput improvements (Table~\ref{tab:bf16_improve}) are evaluated on both Configurations 1 and 2, using a fixed context length of 4096 across four models. The evaluation results show that integrating \texttt{bf16} into MemAscend consistently improves throughput across all tested models and configurations. For Configuration 1, the average throughput gain is 27.25\%, peaking at 56.80\% for \texttt{Qwen2.5-7B} with a batch size of 8, and dipping to 13.24\% for \texttt{Llama3.1-8B} with a batch size of 80. Configuration 2 shows an average improvement of 17.08\%, ranging from 9.99\% for \texttt{Qwen2.5-7B} at batch size 20 to 24.21\% for \texttt{Qwen2.5-32B} at batch size 4. These performance gains stem from lower I/O overhead, as shown in Figure~\ref{fig:bf16_io_footprint}, where \texttt{bf16} usage significantly reduces I/O volume per iteration. The improvements are more pronounced at smaller batch sizes, where I/O bottlenecks dominate. In contrast, larger batches increase computation time, masking I/O savings. These results underscore the effectiveness of \texttt{bf16} in optimizing SSD offloading systems.

\begin{table}[!t]
  \centering
  \caption{Throughput improvement from using bf16 optimizer in MemAscend for Configurations 1 (C1) and 2 (C2)}
  \vspace{-0.1in}
  \label{tab:bf16_improve}
  \resizebox{\columnwidth}{!}{%
    \begin{tabular}{|c|c|c|c|}
      \hline
      \textbf{Model Name} & \textbf{Batch (C1 / C2)} & \textbf{C1 Improvement (\%)} & \textbf{C2 Improvement (\%)} \\ \hline
      Llama3.1-8B & 8 / 8   & 28.63 & 19.39 \\ \hline
      Llama3.1-8B & 80 / 20 & 13.24 & 11.99 \\ \hline
      Qwen2.5-7B  & 8 / 8   & 56.80 & 18.26 \\ \hline
      Qwen2.5-7B  & 64 / 20 & 22.55 & 9.99  \\ \hline
      Qwen2.5-14B & 8 / 4   & 28.84 & 22.11 \\ \hline
      Qwen2.5-14B & 64 / 16 & 16.73 & 11.80 \\ \hline
      Qwen2.5-32B & 8 / 4   & 33.26 & 24.21 \\ \hline
      Qwen2.5-32B & 48 / 8  & 17.92 & 18.87 \\ \hline
    \end{tabular}%
  }
  \vspace{-0.2in}
\end{table}

\paragraph{bf16 Mixed-Precision Training}

This section focuses on mixed-precision training. Although \texttt{fp16} mixed-precision training has been widely used, \texttt{bf16} mixed-precision training for GPU computation has recently gained popularity as an alternative approach. Both methods have their advantages and disadvantages. \texttt{fp16} offers higher precision and supports a wider range of GPUs but requires gradient overflow checks, whereas our method minimizes the overhead associated with this process. In contrast, \texttt{bf16} provides lower precision and requires newer GPU architectures (post-Ampere), but offers greater stability. To demonstrate the memory efficiency of our MemAscend approach in mixed-precision training of \texttt{bf16}, refer to Figure~\ref{fig:bf16_mixed_precision_mem}. Our method achieves an average memory reduction of 25.19\%. However, because \texttt{bf16} has the same data representation range as single-precision float, unlike \texttt{fp16}, it does not require gradient overflow checks and therefore avoids the peak memory overhead associated with such checks; the average memory reduction is lower than that of \texttt{fp16} mixed-precision training, where the average reduction reaches 55.7\%.

\section{Conclusion}\label{sec:conclusion} 

MemAscend re-frames SSD-offloading-based fine-tuning by showing that system memory, rather than GPU capacity, is the decisive constraint on commodity workstations. A detailed analysis uncovers four hidden overheads in the offloading pipeline: buffer-pool fragmentation, over-aligned pinned allocations, overflow-check spikes, and filesystem I/O overhead. MemAscend neutralizes them with the adaptive buffer pool, the alignment-free pinned memory allocation, the fused overflow check mechanism, and the direct NVMe engine for system memory reduction. Together, these techniques cut parameter buffer waste by up to 72.7\%, reduce peak system memory demand by 55.7\%, eliminate the 1.25× overflow check headroom, if users can tolerate a slight loss of precision in optimizer states, MemAscend can improve throughput by up to 56.80\% and 24.21\% on two different hardware configurations when the \texttt{bf16} half-precision optimizer is enabled. The recovered memory headroom immediately translates into support for larger models, longer context windows, and higher batch sizes on unchanged hardware, making full-parameter LLM fine-tuning genuinely accessible to smaller research groups and independent practitioners.


\bibliographystyle{IEEEtran} 
\bibliography{references} 

\vfill

\end{document}